\documentclass[aps,pre,10pt,showpacs,floatfix,onecolumn,nofootinbib,a4paper]{revtex4-2}
\usepackage{amsmath,amssymb,amsthm,mathtools}
\usepackage{mathrsfs}
\usepackage{epsfig,graphicx,xcolor}
\usepackage{verbatim}
\usepackage{paralist}
\usepackage{bm}
\usepackage{float}
\usepackage{dcolumn}
\usepackage{hyperref,bookmark}
\usepackage{url}
\usepackage{enumerate}
\usepackage[normalem]{ulem}
\theoremstyle{definition}
\newtheorem{remark}{Remark}
\newcommand\Real{{\mathord{\rm{I \kern-.22em R}}}}

\newcommand{\free}{\mathscr{F}}
\newcommand{\heat}{\mathscr{Q}}
\newcommand{\dis}{\mathscr{D}}
\newcommand{\ent}{\mathscr{S}}
\newcommand{\ene}{\mathscr{U}}
\newcommand{\work}{\mathscr{W}}
\newcommand{\ray}{\mathscr{R}}
\newcommand{\cycl}{\dot{q}_0}
\newcommand{\real}{\mathbb{R}}
\newcommand{\EGV}[1]{{\color{black}{#1}}}
\newcommand{\PPG}[1]{{\color{black}{#1}}}
\newcommand{\crossedEGV}[1]{{\color{magenta}\sout{#1}}}

\newcommand{\temp}{\vartheta}
\newcommand{\widebar}[1]{\,\overline{\!{#1}}} % overline short italic

\begin{document}
%\latintext

 \title{Analytical Thermodynamics}
 \author{Paolo Podio-Guidugli}
 \email{p.podioguidugli@gmail.com}	
\affiliation{Accademia Nazionale dei Lincei, Via della Lungara 10/230, 00165 Roma, Italy \\and\\ 
	Dipartimento di Matematica, Universit\`a di Roma TorVergata, Via della Ricerca Scientifica 1, 00133 Roma, Italy}
 \author{Epifanio G. Virga}	
 \email{eg.virga@unipv.it}
 \affiliation{Dipartimento di Matematica, Universit\`a di Pavia, Via Ferrata 5, 27100 Pavia, Italy}
 \date{\today}
%\author{ and Epifanio G. Virga} 
%\affiliation{P. Podio-Guidugli \at Dipartimento di Ingegneria Civile, Universit\`a di Roma TorVergata, Viale Politecnico 1, 00133 Roma, Italy \\
%	\email{p.podioguidugli@gmail.com}           %  \\
%	\and
%	E. G. Virga \at Dipartimento di Matematica, Universit\`a di Pavia, Via Ferrata 5, 27100 Pavia, Italy\\
%	\email{eg.virga@unipv.it}    
%}
%\date{Received: xx / Accepted: xx}

%\documentclass[11pt]{article}

\begin{abstract}
	This paper proposes a theory that bridges  classical analytical mechanics and nonequilibrium thermodynamics. Its intent is to derive the evolution equations of a system from a stationarity principle for a suitably augmented Lagrangian action.  This aim is attained for homogeneous systems, described by a finite number of state variables depending on time only. In particular, it is shown that away from equilibrium free energy and entropy are independent constitutive functions.
%	\keywords{Nonequilibrium Thermodynamics\and Analytical Mechanics  \and Homogeneous Systems}
%	\PACS{	45.20.-d   \and 45.20.Jj \and 05.70.-a \and 05.70.Ln }
%	\subclass{70H30  \and 80A05 \and 80M30}
\end{abstract} 
\maketitle

\section{Introduction}\label{sec:into}
We find it appropriate to begin by putting briefly our theory   into historical perspective; this we do in Sect.~\ref{sub1} here below. Next, in Sect.~\ref{sec:sub2}, we consider some variational formulations of thermodynamics. Finally, in Sect.~\ref{sub2}, we introduce the stationarity principle of our choice and summarize the contents of our paper. 

\subsection{Primordial Variational Principles}\label{sub1}
Variational principles  have a long, fascinating history. Perhaps, the first variational principle was put forward more than 2000 years ago by Hero of Alexandria, who postulated that a light ray bouncing on a flat mirror follows the shortest path connecting two fixed points above the mirror. The law of reflection, stating that incoming and outgoing rays make the same angle with the normal to the mirror, then follows as a consequence.\footnote{At first sight, this start might not seem related to the matter at hand, but the reader will soon discover that it really is.} 
If light propagates at constant speed in a medium, then Hero's principle easily converts into the principle of least time, formulated by Fermat in 1662. Elaborating on this, Fermat was able to derive Snell's law of refraction. Such an achievement, however, did not suffice to make Fermat's variational principle universally accepted. There were rival hypotheses, advanced by Descartes, Newton and Leibniz, all equally capable of accounting for Snell's law, as narrated in Chapt.\,1 of Lemons' book \cite{lemons:perfect}. It was not just a matter of rivalry between opposing hypotheses (and schools): there were perplexing counterexamples too. In particular, Descartes' followers showed that light impinging on a \emph{curved}  mirror may well follow the \emph{longest} path joining two given points (see, p.\,8 of \cite{lemons:perfect}), instead of the shortest one.\footnote{The same counterexample shows that in a degenerate case the path of light may neither be the shortest nor the longest.} 
This incident simply teaches us that variational principles significant to physics should prudently be formulated as \emph{stationarity} principles, thus renoucing any teleological implications with which they have too often been impregnated. As Lemons wittily puts it, ``Apparently, nature is extravagant, as well as economical'' \cite[p.\,8]{lemons:perfect}.

\subsection{Thermodynamic Variational Principles}\label{sec:sub2}
Thermodynamics has not been exempt from variational formulations, especially in  its branch concerned with the evolution of states, often called \emph{nonequilibrium} thermodynamics. Here, coherently with the true stance of Fermat's principle, we shall mainly interpret these principles as stationarity requirements.

As appropriately summarized in \cite{gay-balmaz:lagrangian_review},  there are different types of  variational principles for nonequilibrium thermodynamics. There are
principles governing the rate of \emph{entropy production}, with the intent of describing solely the evolution of the \emph{irreversible} processes involved. 
The principle of \emph{least dissipation} of energy, formulated by Onsager in a series of papers \cite{onsager:reciprocal_I,onsager:reciprocal_II,onsager:fluctuations,machlup:fluctuations}, introduced a functional whose minimum determines the transition probability for the system from one state to another. This theory allowed to compute the probability density of thermodynamic variables, but it was confined to linear irreversible processes and Gaussian fluctuations. An extension to nonlinear processes can be found in \cite{serdyukov:extension}.

More recently, Onsager's principle has been the object of further reinterpretation and extension to soft matter systems in the work of Doi \cite{doi:variational,doi:onsager,doi:onsager_tool,doi:application,doi:onsager_polymer} (see also \cite{wang:onsager} for an application to active soft matter).
In the same vein could be placed Gyarmati's principle, from which, building on earlier works of Verh\'as and Vojta, one can derive the canonical equations of thermodynamics (see, in particular, Sect.\,VI.8 of \cite{gyarmati:nonequilibrium}).\footnote{The reader could also profit from consulting the review \cite{ichiyanagi:variational_1994} on this subject.}

In the vast landscape of variational thermodynamic theories, a special role is played  by the work of Biot \cite{biot:variational,biot:variational_book,biot:virtual}. What  is relevant to our developments is Biot's use of a d'Alembertian principle, which may also be described as a principle of minimal \emph{reduced} dissipation, as proposed in \cite{sonnet:dissipative} (see, in particular, Sect.\,2.2.3), where it is used to derive the dynamical equations for dissipative ordered fluids. Specifically, the dissipation is delivered by a Rayleighian function $\ray$ and the quantity to be minimized is $\widetilde{\ray}:=\ray-\work$, where $\work$ is the total working.\footnote{Equivalently, as in Biot's original works, $\ray$ could be minimized under the constraint that both generalized forces and total working are kept fixed (see \cite[p.\,119]{sonnet:dissipative}). In such a formulation, this principle is also known as the principle of minimum \emph{constrained} dissipation.}

There are also principles governing the \emph{whole thermodynamic} evolution away from equilibrium,\footnote{Or towards equilibrium, in the absence of external agencies.} with the intent of describing \emph{all} processes involved, reversible and irreversible alike.

Although not strictly variational in nature, the formalism of \emph{dissipative Poisson brackets}, as expounded \EGV{in general terms} for example in \cite{grmela:dynamics,ottinger:dynamics,baldiotti:hamiltonian},\footnote{\EGV{Going backwards in time, these papers were preceded, essentially in the same line of thought, by \cite{grmela:bracket}, while the origin of the method can be retraced in an early paper by Kaufman and Morrison~\cite{kaufman:algebraic}, framed in the context of the quasilinear equations of plasma physics. This  paper was later followed by \cite{kaufman:dissipative,morrison:bracket,morrison:paradigm}. More recent applications of the method can be found in the specific fields of multiphase fluids \cite{eldred:single} and dissipative magnetohydrodynamics\cite{coquinot:general}.}} comes closer to the theme of our paper, which is concerned with deriving the equations that govern the thermodynamic evolution away from equilibrium with the methods of classical analytical dynamics.\footnote{\EGV{A similar objective was recently pursued in \cite{cendra:elementary}, but \emph{without} accounting for the \emph{second} law of thermodynamics.}} 

\subsection{Our Approach} \label{sub2}
The natural antecedent of our work is that of Gay-Balmaz and Yoshimura \cite{gay-balmaz:lagrangian_I,gay-balmaz:lagrangian_II,gay-balmaz:variational} (see, in particular, the review \cite{gay-balmaz:lagrangian_review}). In this theoretical approach, a Hamilton's action functional is required to be stationary under a \emph{nonholonomic} constraint for the \emph{thermal} variables of the system.\footnote{Despite the novelty of this approach, studies on the Lagrange-Hamilton formalism for nonequilibrium thermodynamics had  appeared before  in the literature: we mention, for example, \cite{gambar:hamilton} and refer the reader to the discussion in \cite{van:structure}.} Such a constraint, which is linear in the generalized velocities, incorporates the entropy production. The demand of stationarity for the Hamiltonian functional is phrased in a fashion akin to the Lagrange-d'Alembert principle, that is, by requiring stationarity of the action under the linear variational constraint generated by replacing the generalized velocity with variations of the generalized variables.\footnote{As remarked in \EGV{\cite[p.\,44]{arnold:mathematical}}, the equations of motion generated by this principle differ from those (also called \emph{vakonomic}) that are generated by requiring stationarity of the action under the nonholonomic constraint meant to restrict accessibility in phase space. \EGV{Vakonomic dynamics was developed by V.~V. Kozlov in a series of six papers published in the 1980's (referenced as [329] in \cite[p.\,490]{arnold:mathematical}).}}

%\paragraph
Our approach to analytical thermodynamics is different. We  seek compatibility between the  evolution equations of a system with a finite number of state variables and the laws of thermodynamics as formulated within Truesdell's theoretical framework \cite{truesdell:rational}. \EGV{In our view, this represents a novel approach to \emph{nonequilibrium} thermodynamics, in which the evolution out of equilibrium of a system is governed within a consistent Lagrangian formalism.} The case of homogeneous systems is illustrated in Sect.~\ref{sec:homo}. 
Thermal variables (here, for simplicity, a single one) 
are envisaged as \emph{cyclic}, in accord with  Helmholtz's mechanical interpretation of thermodynamics summarized in Sect.~\ref{HMIT}. Their time derivatives enter the Lagrangian as a macroscopic manifestation of microscopic motions too fast to be detected. Sect.~\ref{sec:Rayleigh} is devoted to illustrate the role played by Rayleighian potentials in extending Hamilton's functional to dissipative systems. Sect.~\ref{mod} is a summary of elementary analytical mechanics, which serves the dual purpose of making our exposition self-contained and fix our notation. At its opposite end, our paper is closed by Sect.~\ref{sec:conclusion}, where we outline our conclusions. 
%\newpage
\section{A Modicum of Analytical Mechanics}\label{mod}
Lagrange was first to see a serious limitation of Newton's motion laws, namely, to be expressed in terms of the current \emph{positional coordinates} of the mass points in a finite system, as observed against a fixed spacetime background. He was  able to formulate the evolution laws of any system with finite degrees of freedom in terms of a stationarity condition for a system-specific \emph{action functional} depending on the \emph{generalised coordinates} of the objects in the system and on the time rates of those coordinates. 

Hamilton quickly followed up with an even more general formulation, in which an %equally 
orderly list of \emph{kinetic momenta} replaces for the list of Lagrange's rates of generalised coordinates.  
Analytical Mechanics (AM) as we reckon it today is in Hamiltonian form; its general purpose is to describe the deterministic and nondissipative evolution of a system with a finite number of degrees of freedom. 
%

%\markboth{P. Podio-Guidugli}{Chapter \ref{anth}. Analytic Thermodynamics \draft}
%
\subsection{Lagrangians}\label{16.1}
Classically, a \emph{Lagrangian function}--briefly, a \emph{Lagrangian}--is a mapping
\begin{equation}\label{L}
	L=L(q_1,q_2,\ldots,q_n; \dot q_1, \dot q_2,\ldots, \dot q_n;t)
\end{equation}
depending on  a finite list of configuration parameters $q_i$, the  \emph{Lagrangian coordinates}, their time derivatives $\dot q_i$, and time $t$. 

The \emph{Lagrangian equations of motion} of the mechanical system described by $L$ are
\begin{equation}\label{lagra1}
	\frac{\partial L}{\partial q_i}-\frac{d\,}{dt}\left(\frac{\partial L}{\partial \dot q_i}\right)=0 \quad(i=1,2,\ldots,n);
\end{equation}
they express the \emph{stationarity conditions} of the \emph{Lagrangian action functional}
\begin{equation}\label{actionL}
	{\mathcal A}_L\{q\}=\int_{t_1}^{t_2}L(q_1,q_2,\ldots,q_n; \dot q_1, \dot q_2,\ldots, \dot q_n;t)\,dt
\end{equation}
at a system's path 
\begin{equation}\label{Lpath}
	t\mapsto q(t)=(q_1(t),q_2(t),\ldots,q_n(t)),
\end{equation}
for whatever variation $\delta q$  keeping the path ends $q(t_1)$ and $q(t_2)$ fixed. Precisely, equations \eqref{lagra1} are arrived at by putting to null
\[
\frac{d\,}{d\varepsilon}\big({\mathcal A}_L\{q+\varepsilon \delta q\}\big)
\]
at $\varepsilon=0$ and exploiting the quantification with respect to  admissible variations. 
Equations \eqref{lagra1} integrate a second-order system of $n$ ODEs for $q(\cdot)$.   Under the standard assumption that
\begin{equation}\label{detlag}
	\det [L_{ij}]\neq 0,\quad L_{ij}:=\frac{\partial^2 L}{\partial \dot q_i\partial\dot q_j}\,,
\end{equation}
\EGV{identifying  \emph{nondegenerate} Lagrangians,}
this system may be put in the \emph{normal form}
\begin{equation}\label{nf}
	\ddot q_i=f_i(q,\dot q,t) \quad(i=1,2,\ldots,n),
\end{equation}
and solved uniquely for given initial conditions
\begin{equation}\label{ic}
	q(t_0)=q_0,\quad \dot q(t_0)=\dot q_0.
\end{equation}

The (\emph{total}) \emph{energy} associated with a Lagrangian $L$ is
\begin{equation}\label{toten}
	E(q,\dot q,t):=\partial_{\dot q}L(q,\dot q,t)\cdot\dot q - L(q,\dot q, t),\quad\textrm{where}\quad \partial_{\dot q}L\cdot\dot q=\sum_{i=1}^n \frac{\partial L}{\partial \dot q_i}\dot q_i ,
\end{equation}
\EGV{a definition valid irrespective of the validity of \eqref{detlag}}.
This terminology is justified by thinking of \EGV{the case that we  call  \emph{dynamical}}, that is, the familiar instance when a given Lagrangian is split as follows:\footnote{\label{foot:wittaker}\EGV{Calling this case \emph{dynamical} is typical of the Italian School of Mechanics (see, for example, \cite[p.\,247]{levi_civita:lezioni}). Along similar lines, in the British tradition, (see, for example, \cite[p.\,57]{whittaker:treatise}), the word \emph{natural} is used when the Lagrangian contains only terms of degree $2$ or $0$ in the velocities.}} 
\begin{equation}\label{lagsplit}
	L=K-U
\end{equation}
into \emph{kinetic energy} 
\begin{equation}\label{cinen}
	K=K(q,\dot q,t), \quad\textrm{with}\quad 2K=\partial_{\dot q} K\cdot \dot q,
\end{equation}
and \emph{potential energy} 
\begin{equation}\label{U}
	U=U(q,t). 
\end{equation}
In that case, 
\begin{equation}\label{emaiu}
	E=\partial_{\dot q}K(q,\dot q,t)\cdot\dot q - K(q,\dot q, t)+U(q,t)=K(q,\dot q,t)+U(q,t),
\end{equation}
and kinetic plus potential energies make up the total energy. Note that \eqref{lagsplit} and \eqref{U} imply that
\begin{equation}\label{detkin}
	L_{ij}=\frac{\partial^2 K}{\partial \dot q_i\partial\dot q_j}=:a_{ij}\,;
\end{equation}
moreover, it follows from $\eqref{cinen}_2$ that $\,\partial_{\dot q_i}K=a_{il}\dot q_l$,
and hence that the kinetic energy is a quadratic form, ruled by the symmetric matrix $a_{ij}$:
\begin{equation}
	\label{eq:kinetic_lego}
	K=\frac{1}{2}\sum_{i,l=1}^na_{il}\dot q_i\dot q_l\,;
\end{equation}
the standard assumption that the kinetic energy be positive is more than enough to guarantee that $\eqref{detlag}_1$ hold.

Interestingly, \emph{if a Lagrangian function does not depend explicitly on time, then the associated energy is a motion constant (and conversely)}. In fact, from definition (\ref{toten}) we have that
\[
\frac{dE}{dt}=\frac{d\,}{dt}\big(\partial_{\dot q}L\big)\cdot\dot q+\partial_{\dot q}L\cdot\ddot q-\partial_qL\cdot\dot q-\partial_{\dot q}L\cdot\ddot q-\partial_tL =
\Big(\frac{d\,}{dt}\big(\partial_{\dot q}L\big)-\partial_qL\Big)\cdot\dot q-\partial_tL,
\]
whence
\begin{equation}\label{ddtE}
	\frac{dE}{dt} = -\,\partial_tL
\end{equation}
along all solutions of (\ref{lagra1}); the announced result follows directly from this motion identity.
\subsection{Hamiltonians}\label{hamil}
Hamilton proposed to encode all the properties of a mechanical system in a  \emph{Hamiltonian function}
\begin{equation}\label{hamm}
	H=H(q_1,q_2,\ldots,q_n; p_1, p_2,\ldots,  p_n;t),
\end{equation}
a mapping generally depending, in addition to time and a finite list $q$ of configurational parameters, not on the time rates of the latter but on a list  $p=(p_1,p_2,\ldots,p_n)\,$ of  \emph{kinetic momenta}.\footnote{\PPG{Kinetic} momenta, \EGV{as they are called in the Italian tradition (see, for example, \cite[p.\,246]{levi_civita:lezioni}),} were introduced by Hamilton, who called them \emph{canonical}, an adjective which \EGV{is used in more than one sense in the literature.}} He wrote the system's \emph{Hamiltonian equations of motion} as follows:

\begin{equation}\label{hamit}
	\dot q_i=\frac{\partial H}{\partial p_i},\quad \dot p_i=-\frac{\partial H}{\partial q_i} \quad(i=1,2,\ldots,n),
\end{equation}
a first-order system of $2n$ ODEs for the pair of functions of time $(q(\cdot),p(\cdot))$ delivering the time evolution of the \emph{Hamiltonian coordinates} $(q,p)$ of the mechanical system at hand. Just as the Lagrangian equations for the action functional \eqref{actionL}, the Hamiltonian equations of motion (\ref{hamit}) can be shown to be the stationarity conditions at a path $t\mapsto (q(t),p(t))$ of the \emph{Hamiltonian action functional}
\begin{equation}\label{actionH}
	{\mathcal A}_H\{q,p\}=\int_{t_1}^{t_2}\!\!\left(p\cdot \dot q-H(q, p,t)\right)\,dt,\quad p\cdot \dot q=\sum_{i=1}^n p_i\dot q_i,
\end{equation}
for all path variations keeping $q(t_1)$ and $q(t_2)$ fixed.\footnote{\EGV{The functional $\mathcal{A}_H$ is also called the \emph{phase space action} (see, for example, \cite[p.\,37]{arnold:mathematical}).}}

\begin{remark}
	Both Lagrange's and Hamilton's stationarity principles are formulated under identical restrictions on admissible path variations at the boundaries of the considered time interval. Consequently, \emph{stricto sensu} both Lagrangian and Hamiltonian equations of motions \eqref{lagra1} and \eqref{hamit} may solve a boundary-value problem in time for assigned values of $q(t_1)$ and $q(t_2)$; they are, however, commonly used to formulate time-evolution problems for given initial data \EGV{(see also \cite{Galley}).}
\end{remark}

%\noindent{\bf{Remark.}}
%With the use of the notion of \emph{Poisson bracket}
%%
%\begin{equation}\label{Pbra}
% \{A,B\}:=\frac{ A}{ a}\cdot\frac{ B}{ b}-\frac{ A}{ b}\cdot\frac{ B}{ a}
% \end{equation}
% %
% of two functions $A=A(a,b)$ and $B=B(a,b)$, and on setting $z=(q,p)$, the Hamiltonian equations for the motion $t\rightarrow z=\zeta(t)$, can be written in the compact form
%%
%\begin{equation}\label{PHam}
%\dot z=\{z,H\}.
%\end{equation}
%%
%
%

%On introducing the \emph{kinetic momenta}  conjugated with the time rates of the Lagrangian coordinates, namely,
%%
%\begin{equation}\label{mom}
%p_i:=\frac{\partial L}{\partial \dot q_i}\quad(i=1,2,\ldots,n),
%\end{equation}
%%
%the Lagrangian equations of motion (\ref{lagra1}) can be written as follows:
%%
%\begin{equation}\label{lagra22}
%\dot p_i=\frac{\partial L}{\partial q_i} \quad(i=1,2,\ldots,n).
%\end{equation}
%%
%In so doing, (\ref{lagra1}), a second-order system of $n$ equations, is transformed into the first-order system of $2n$ equations consisting of (\ref{mom}) and (\ref{lagra22}). There are countless ways of performing such a transformation; the most efficient one--the so-called \emph{canonical transformation}--was introduced by Hamilton. A preparatory idea is to regard each of relations \eqref{mom}--the $i$-th, say--as an implicit equation in the unknowns $(p_i, q,\dot q,t)$, that, in view of assumption (\ref{detlag}), can be solved for $\dot q_i$:
%%
%\begin{equation}\label{dotqp}
%\dot q_i=g_i(q,p,t)
%\end{equation}
%%
%(here, just as we did with $q$, $p\equiv(p_1,p_2,\ldots,p_n)\,$). 

%
\subsection{Lagrangians vs. Hamiltonians}\label{LvsH}
A question comes to mind: given a mechanical system, are its Lagrangian and Hamiltonian descriptions interchangeable?
%when is it that a Lagrangian $L$ and a Hamiltonian $H$ describe the same mechanical system? 
With a view toward answering, we develop a couple of preparatory considerations.

Firstly, for $L$ a Lagrangian function as in \eqref{L}  we introduce the kinetic momenta  conjugated with the time rates of the Lagrangian coordinates, namely,
\begin{equation}\label{mom}
p_i:=\frac{\partial L}{\partial \dot q_i}\quad(i=1,2,\ldots,n).
\end{equation}
With this notion at hand, the Lagrangian equations of motion (\ref{lagra1}) can be written as follows:
\begin{equation}\label{lagra22}
\dot p_i=\frac{\partial L}{\partial q_i} \quad(i=1,2,\ldots,n).
\end{equation}
We see that (\ref{lagra1}), a second-order system of $n$ equations, is transformed into the first-order system of $2n$ equations consisting of (\ref{mom}) and (\ref{lagra22}). 
%There are countless ways of performing such a transformation; we anticipate from Section \ref{simplref} that the most efficient format the motion equations of a system with a finite number of degrees of freedom can be given is a so-called \emph{canonical} Hamiltonian formulation.

As a second preparatory step, we regard each of relations \eqref{mom}--the $i$-th, say--as an implicit equation in the unknowns $(p_i, q,\dot q,t)$, an equation that, in view of assumption (\ref{detlag}), can be solved for $\dot q_i$:\footnote{
\EGV{More precisely, assumption \eqref{detlag} only guarantees the local invertibility of \eqref{mom}. As shown in \cite[Sect.\,3.11]{gallavotti:elements}, for \eqref{dotqp} to be valid in a whole neighborhood, we must strengthen \eqref{detlag} by assuming that the mapping
	$$
	\dot{q}\mapsto\frac{\partial L}{\partial \dot{q}}
	$$
	be a diffeomorphism. In \cite{gallavotti:elements}, Lagrangians with such a property are also called \emph{regular}. Dirac  considered in \cite{dirac:generalized} a more general form of Hamiltonian dynamics which can also be applied  when momenta are \emph{not} independent functions of (generalized) velocities.}}
\begin{equation}\label{dotqp}
\dot q_i=g_i(q,p,t);
\end{equation}
for later convenience, we rewrite these $n$ relations in the compact form
\begin{equation}\label{compatta}
\dot q=g(q,p,t).
\end{equation}
We are now ready to answer the question we posed in the affirmative. 

\paragraph{From $L$ to $H$:} The Hamiltonian function corresponding to a given Lagrangian $L$ is constructed by setting
\begin{equation}\label{H}
H(q,p,t):=p\cdot\dot q-L(q,\dot q,t),
\end{equation}
so as to let the integrands of the associated Lagrangian and Hamiltonian action functionals take the same values along the same path.\footnote{It is straightforward to check that such an $H$ does not depend on $\dot q$. Indeed, due to definition (\ref{mom}), 
%of the kinetic momenta $p$ conjugated to the Lagrangian rates $\dot q$,
%
\[
\frac{\partial H}{\partial \dot q_i}= p_i-\frac{\partial L}{\partial \dot q_i}=0.
\]
} Combining (\ref{H}) with (\ref{dotqp}) and \eqref{compatta}, we have that
\begin{equation}\label{HH}
H(q,p,t)=p\cdot g(q,p,t) - L(q,g(q,p,t), t).
\end{equation}
Now, by taking $H$ as given by \eqref{H} in the Hamiltonian equations of motion \eqref{hamit}, we quickly see that the first of \eqref{hamit} reduces to an identity, while the second takes the form \eqref{lagra1}. All in all, equations \eqref{hamit} may be regarded as the Hamiltonian form of the Lagrangian equations \eqref{lagra1}. 

\paragraph{From $H$ to $L$:} The Lagrangian corresponding to a given Hamiltonian $H$ obtains by postulating that the first of (\ref{hamit}) delivers the time rates of the Lagrangian coordinates conjugated to the kinetic momenta,  and by setting
\begin{equation}\label{LH}
L(q,\dot q,t):=p\cdot\dot q -H(q,p,t).
\end{equation}
Such an $L$ is independent of $p$ \EGV{by  \eqref{hamit}$_1$};  moreover, by (\ref{mom}),
\begin{equation}\label{pq}
p_i=h_i(q,\dot q,t),\quad \text{or rather}\quad p=h(q,\dot q,t),
\end{equation}
whence the following form for (\ref{LH}):
\begin{equation}\label{LHL}
L(q,\dot q,t)= h(q,\dot q,t)\cdot\dot q-H(q,h(q,\dot q,t),t).
\end{equation}

We observe that 
\begin{equation}\label{hamcons}
\dot H=\partial_q H\cdot\dot q+\partial_p H\cdot\dot p+\partial_t H=\partial_t H
\end{equation}
along any motion. Hence, \emph{system's Hamiltonian is a motion constant iff it does not depend on time explicitly}. Relations (\ref{H}) and (\ref{LH}) show that either both a Hamiltonian and the corresponding Lagrangian depend  on time explicitly or neither does. 

We also observe that relations \eqref{HH} and \eqref{LHL} demonstrate that, given a mechanical system, its Hamiltonian is the \emph{Legendre transform} of its Lagrangian, and conversely. The involutory nature of the Legendre transform is made patent by either  \eqref{H} or \eqref{LH}, both of which imply that
\[
L(q,\dot q,t)+H(q,p,t)=p\cdot\dot q.
\]
\begin{comment} 
(the elementary properties of Legendre transforms are listed in Sect. \ref{appendix}). 
\end{comment}
\subsection{Hamiltonian Handling of the Dynamical Case}

%\vskip 4pt
%
%\noindent{\bf{Remark 3.}}\label{16.1}
% In the dynamical case, 
We begin with showing that, in this case,
\begin{equation}\label{haccae}
H=K+U=E,
\end{equation}
a motion constant: in view of \eqref{lagsplit}-\eqref{U} and \eqref{mom}, 
\[
K=\frac{1}{2}\,p\cdot\dot q\,;
\]
with this, \eqref{haccae} follows from \eqref{H}, again \eqref{lagsplit}, and \eqref{emaiu}. 

Moreover, on setting
\begin{equation}
p_i=m_i\dot q_i%\;\;(\textrm{no sum on index}\; i),
\end{equation}
with $m_i$, a positive constant, the \emph{mass} associated with the Lagrangian coordinate $q_i$, we have that
\begin{equation}\label{dinca}
H(q,p)=K(p)+U(q)\,,\quad\textrm{with}\quad K(p)=\frac{1}{2}\sum_{i=1}^n m_i^{-1}p_i p_i.
\end{equation}
Accordingly, the first of Hamiltonian motion equations  (\ref{hamit}) becomes an identity, the second takes a familiar Newtonian form: 
\begin{equation}\label{hamin}
m_i\ddot q_i=\big(\dot p_i=-\partial_{q_i} H\big)=-\partial_{q_i} U=: f_i \quad(i=1,2,\ldots,n),
\end{equation}
with $f_i$ the \emph{force} acting on the system whenever the value of its $i-$th Lagrangian coordinate changes. On expressing the kinetic energy as
\[  
K=\frac{1}{2}\sum_{i=1}^n m_i \dot q_i \dot q_i
\]
and on setting
\begin{equation}\label{forz}
f_K:=- \frac{d}{dt}\big(\partial_{\dot q}K \big)
\;\;\,\text{and}\;\;\,f_U:= -\partial_q U
\end{equation}
for, respectively, the \emph{inertial force} and the \emph{potential force},\footnote{When restricting attention to potential energies that do not depend explicitly on time, the standard qualifier  for $f_U$ is  \emph{conservative}.} we can reformulate \eqref{hamin} under form of the \emph{force balance}
\begin{equation}\label{forzbal}
f_K+f_U=0,
\end{equation}
a relation which is quickly shown equivalent to
\begin{equation}\label{Hcost}
\dot H=0.
\end{equation}

Finally,  we let $U_A:=U(q_A)$,  $K_A:=K(p_A)$ and call 
\begin{equation}\label{workperf}
W_{AB}:=U_B-U_A,\;\;
\end{equation}
the work performed on the system when it evolves from the state $(q_A,p_A)$ to the state $(q_B,p_B)$. Then,  the fact that the system's Hamiltonian is a motion constant implies the so-called \emph{work-and-kinetic energy theorem} 
\begin{equation}\label{worken}
W_{AB}=K_A-K_B
\end{equation}
according to which an increase (decrease) in potential energy is accompanied by a decrease (increase) in kinetic energy.
% \clearpage
\section{Rayleigh-Lagrange Dynamics}\label{sec:Rayleigh}
The dynamical case is of special importance in classical Statistical Mechanics (SM), where the typical system consists of a large number of constant-mass points, whose positions in space are the system's Lagrangian coordinates, acted upon by conservative forces.  That theory 
stops short from handling systems which are not Hamiltonian, because they are acted upon by forces that (like e.g. frictional forces) cannot be derived from a potential function.
%, something that a Newtonian approach instead permits. 
However, 
%as we shall see in Section \ref{nonH}, 
the reach of SM can be so extended  as to handle \emph{non-Hamiltonian systems} whose evolution manifests energy dissipation (see \cite{Tu99,Tu,TuCic}). 

In AM, handling dissipative forces was first made possible in 1871, by a change of format due to Rayleigh \cite{strutt,rayle,raycoll}.

Let us introduce Rayleigh's \emph{dissipation potential}  
%and \emph{dissipation function}
$R=R(q,\dot q)$, with $R(q,\cdot)$ a positive semidefinite quadratic form for all $q$,
% %
and the associated \emph{dissipative force}
\begin{equation}\label{ray}
f_R:=-\partial_{\dot q}R.
\end{equation}
% %
The corresponding \emph{Rayleigh-Lagrange dynamics}   is ruled by a force balance that generalizes \eqref{forzbal}, namely,
\begin{equation}\label{forzbalray}
f_K+f_U+f_R=0.
\end{equation}
%
% the stationarity condition of the \emph{Rayleighian-Lagrangian action}
%%
%\begin{equation}\label{actionL}
%{\mathcal A}_{RL}\{q,\dot q\}=\int_{t_0}^{t_1}\!\!\big(L(q,\dot q;t)+R(q,\dot q)\big)\,dt\,.
%\end{equation}
%%
%at a system's path 
%%
%\begin{equation}\label{Lpath}
%t\rightarrow q(t)\equiv(q_1(t),q_2(t),\ldots,q_n(t)),
%\end{equation}
%%
%for whatever variation $\delta q$  keeping the path ends $q(t_0)$ and $q(t_1)$ fixed.
It is the matter of a simple computation to show that  under the present circumstances \eqref{Hcost} is replaced by
\begin{equation}\label{raylag}
\dot H-D=0,
\end{equation}
with
\begin{equation}\label{DeR}
D(q,\dot q):=\dot q\cdot\partial_{\dot q} R.
\end{equation}

As suggested in \cite{virga:rayleigh}, the roles of $R$ and $D$ may be reversed, in that  relation \eqref{DeR} may be regarded as a partial differential equation to be solved for  the Rayleigh's potential $R$ conveying  the information about a system's dissipative features embodied in a physically plausible choice of function $D$. It is proposed in \cite{virga:rayleigh} that $D(q,\cdot)$ be chosen positive semidefinite\footnote{\EGV{That is, such that $D(q,\cdot)\geqq0$ for all $q$.}} (but
not necessarily quadratic, as instead is commonly done) and  vanishing identically at $\dot q=0$. 

%{\color{blue}Having recourse to this stationarity condition is indeed the operational definition of \emph{Rayleigh's principle of minimal dissipation}. }

The so-called \emph{overdamped regime} comes about when the inertial force is negligible with respect to the dissipative force and  a system's time evolution is such that
\begin{equation}
\dot H\simeq \dot U.
\end{equation}
If, in addition, it is assumed that
\begin{equation}\label{zeta}
R(q,\dot q)=\frac{1}{2}\,Z(q)\cdot\dot q\otimes\dot q\geq 0,\;\; Z=Z^T,
\end{equation}
then the force balance \eqref{raylag} reduces to the \emph{kinetic equation}
\begin{equation}\label{kineq1}
\partial_q U+Z \dot q=0,
\end{equation}
a condition that is quickly seen to be necessary for the stationarity at any fixed $q$ of the \emph{Rayleighian}
\begin{equation}\label{rayfu}
\mathcal{R}(q,\dot q):=\partial_q U\cdot\dot q+R(q,\dot q).\end{equation}

In case matrix $Z$ in \eqref{zeta} is taken invertible, the kinetic equation \eqref{kineq1} can be written as
\begin{equation}\label{kineq2t}
\dot q=-Z^{-1}\partial_q U;
\end{equation}
provided $U$ is interpreted as the system's \emph{free energy}, \eqref{kineq2t} may be regarded as the stationarity condition of an \emph{Onsager-like} variational principle, in the wake of \cite{onsager:reciprocal_I}, as recently reinterpreted in much soft matter literature.\footnote{We read in \cite{doi:onsager_tool}: ``The Onsager principle is an extension of Rayleigh's
principle of the least energy dissipation in Stokesian
hydrodynamics."}
\EGV{
\begin{remark}
	Rayleigh-Lagrange dynamics for dissipative systems have a different standing than ordinary Lagrangian dynamics. The \PPG{latter} can be given a variational formulation in the classical form of stationary of the action functional in \eqref{actionL}, while the \PPG{former} can be given a variational formulation only in a d'Alembert fashion by requiring the stationarity of an action \emph{augmented} by the addition of the \emph{virtual} work expended by dissipative forces. The form appropriate to our development\PPG{s} of \PPG{the stationarity of such an augmented action} will be written in \eqref{eq:Hamilton_principle_augmented}.
\end{remark}
}

\section{Mechanical Interpretations of Thermodynamics}\label{HMIT}
Helmholtz, who was working before Boltzmann's creation of a statistical approach to physics, pursued a purely mechanical interpretation of the objects of thermodynamics and  a consistent deduction of the basic laws of that discipline. Interestingly, Boltzmann himself continued along Helmholtz's nonprobabilistic path in his own treatment of periodic Lagrangian systems.

The thermodynamical objects in need of purely mechanical definitions are temperature, heat, entropy, etc. With Helmholtz, we begin by temperature $\vartheta$, that he regarded as a velocity-like variable and accordingly introduced as the time derivative of another object, the \emph{thermal displacement}
\begin{equation}\label{thdis}
\dot\alpha=\vartheta,
\end{equation}
whose physical interpretation was \PPG{at the time} rather mysterious.\footnote{Perhaps, not anymore, see \cite{PPGThD}. For a demonstration of the role of thermal displacement in a virtual-power formulation of thermomechanics, see \cite{PPGlv} and \cite{SISSA1}. The reader is also referred to \cite{green:re-examination} for  a different theory building on this notion, and to \cite{gurtin:general}, where the theory in \cite{green:re-examination} is to some extent anticipated.} The formal role Helmholtz assigned to thermal displacement was that of a \emph{cyclic coordinate}, that is, a coordinate appearing in the Lagrangian of a mechanical system only through its time derivative, whose momentum is hence a motion constant \cite{H1,H2} (see Lanczos \cite{L}, pp. 125 ff.).  
With L. de Broglie \cite{Bro1,Bro2,Bro3}, we call it a `fast' variable, implying that only its time derivative is observable at the macroscopic time scale and is therefore included in the list of Lagrangian coordinates. Accordingly, we call `slow' the ordinary Lagrangian coordinates, which are observable and have observable time derivative.\footnote{In \cite{wang:onsager}, when ``the general steps for applying OVP [Onsager Variational Principle] to the dynamics of active soft matter" are listed, the first step is ``[t]o choose a set of coarse-grained slow variables  \ldots to describe the time evolution of the macroscopic state of the system."}
% \footnote{
%After Helmholtz's original proposal, a notion of thermal displacement found manifold uses in the physical literature: in addition to the works of v. Laue \cite{vL}, v. Dantzig \cite{vD} (who apparently was responsible for changing v. Laue's `therma\underline{c}y' into `therma\underline{s}y') and de Broglie \cite{Bro2,Bro3}, the thermasy concept enters, e.g., in the high-energy physics and cosmology issues treated by Roberts \cite{R1,R2} and by Kuzminchev \& Kuzminchev \cite{KK1,KK2}.
%At the moment of this writing, a thorough discussion of this notion is still wanted; suffice it to quote here some  papers where thermal displacement plays a role, in developing a Hamiltonian structure for either relativistic perfect fluids \cite{BMW,KSG,Br} or nonrelativistic dissipative materials  \cite{M,KM,KM2}, as well as in rethinking certain foundational aspects of thermomechanics with a view toward resolving the infinite wave velocity paradox and modeling second-sound phenomena in solids \cite{GN1,GN3,DM,BS}.} 

\subsection{Helmholtz's Monocyclic Systems}\label{hh1}
Consider a \emph{monocyclic} mechanical system, that is, a mechanical system whose Lagrangian, perhaps modulo a reordering, does not depend on the $n-$th coordinate:
\begin{equation}\label{lagra3}
L=L(q_1,q_2,\ldots,q_{n-1},\mathrlap{q_n}\bigtimes; \dot q_1, \dot q_2,\ldots, \dot q_n;t),
\end{equation}
For such a system, on recalling  \eqref{lagra22}, we have that:
\[
\frac{\partial L}{\partial  q_n}=0\quad\Rightarrow\quad\dot p_n=0\quad \Leftrightarrow\quad p_n=c_n,\;\textrm{a motion constant},
\]
a result that, on recalling \eqref{mom}, we write as
\begin{equation}\label{defcn}
\frac{\partial L}{\partial \dot q_n}-c_n=0,
\end{equation}
\EGV{which makes it clear how $c_n$ is indeed determined by the  initial conditions imposed on the specific motion being studied.}
We regard \eqref{defcn} as an implicit relation among the variables 
%$(q_1,q_2,\ldots,q_{n-1}; \dot q_1, \dot q_2,\ldots, \dot q_n;t)$ 
from which $L$ depends and $c_n$, and assume that it can be cast in normal form:
\begin{equation}\label{mono}
\dot q_n=f(q_1,q_2,\ldots,q_{n-1}; \dot q_1, \dot q_2,\ldots, \dot q_{n-1};c_n;t).
\end{equation}
With the use of (\ref{mono}), the first $(n-1)$ motion equation (\ref{lagra1}) yield the time evolution of the slow coordinates; having this information, we can revert to (\ref{mono}), and obtain the time evolution of the fast coordinate $q_n$ by quadrature.
\begin{remark}\label{rmk:A}
A fast coordinate is not observable, its time rate is. Think of a rigid conductor: its slow Lagrangian coordinates are the six parameters needed to describe its macroscopic rigid-body motion; \emph{temperature} $\vartheta$, a fast coordinate, is observable, but \emph{thermal displacement} $\alpha$, the scalar variable defined in \eqref{thdis}, is not. Just as temperature gives a macroscopic information that we interpret as an account of the velocity fluctuations of a representative collection of body's molecules, the thermal displacement can be regarded as accounting for the nonobservable fluctuations of those molecules about their lattice positions.  
\end{remark}
\begin{remark}\label{rmk:B}
For another example, this time with many fast variables (a \emph{polycyclic} mechanical system, in Helmholtz's terminology),  think of a gas, with the molecule coordinates as the fast variables and the macroscopic parameters that determine the state of the system (pressure, volume, or temperature) as the slow variables.
\end{remark}

\EGV{
Now, following Lanczos~\cite[p.\,126]{L}, we wonder whether we can eliminate the cyclic variable $q_n$ \emph{before} solving the time evolution equations for the non-cyclic variable $q_1,\dots,q_{n-1}$, by reformulating appropriately the parent variational principle. Our aim is to find a Lagrangian $\widebar{L}$ depending \emph{only} on the non-cyclic variables of $L$ (and their time derivatives) so that the stationarity of the action $\mathcal{A}_{\widebar{L}}$ delivers precisely the same evolution equations for the non-cyclic variables of $L$, once use is made in them of \eqref{mono}. 

To this end, 
given a system whose Lagrangian has the form \eqref{lagra3}, we may require that the associated Lagrangian action 
\begin{equation}\label{eq:action_L}
{\mathcal A}_L=\int_{t_1}^{t_2}L(q_1,q_2,\ldots,q_{n-1}; \dot q_1, \dot q_2,\ldots, \dot q_n;t)\,dt
\end{equation}
be varied under the constraint \eqref{mono}:
\begin{equation}\label{eq:variation_on_constraint}
\begin{split}
	\delta\int_{t_1}^{t_2}\big[&L(q_1,q_2,\ldots,q_{n-1}; \dot q_1, \dot q_2,\ldots, \dot q_n;t)\\&+\lambda\big(\dot q_n-f(q_1,q_2,\ldots,q_{n-1}; \dot q_1, \dot q_2,\ldots, \dot q_{n-1};c_n;t)\big)\big]dt=0.
\end{split}
\end{equation}
However, \eqref{mono}, even if incorporated in \eqref{eq:variation_on_constraint} through the Lagrange multiplier $\lambda$ (which at this stage is an unknown function of time), is in general incompatible with having $\delta q_n$ vanishing at both $t=t_1$ and $t=t_2$. It follow\PPG{s} from \eqref{mono} by integration that 
\begin{equation}
\label{eq:integrated_constraint}
\delta q_n(t_2)-\delta q(t_1)=\int_{t_1}^{t_2}\sum_{i=i}^{n-1}\left[\frac{\partial f}{\partial q_i}-\frac{d}{dt}\left(\frac{\partial f}{\partial\dot{q}_i}\right)\right]\delta q_i dt, 
\end{equation}
under the usual assumption that $\delta q_i=0$ at both $t=t_1$ and $t=t_2$ for all $i=1,\dots,n-1$. Thus, taking $\delta q_n(t_1)=0$, the appropriate stationary condition for $\mathcal{A}_L$ in \eqref{eq:action_L} subject to \eqref{mono} is that the variation in \eqref{eq:variation_on_constraint} be proportional through a constant Lagrange multiplier $\mu$ to $\delta q_n(t_2)$ as given by \eqref{eq:integrated_constraint}:
\begin{equation}
\label{eq:constrained_variation}
\delta\int_{t_1}^{t_2}\big(L+\lambda(\dot{q}_n-f)\big)dt-\mu\int_{t_1}^{t_2}\sum_{i=i}^{n-1}\left[\frac{\partial f}{\partial q_i}-\frac{d}{dt}\left(\frac{\partial f}{\partial \dot{q}_i}\right) \right]\delta q_idt=0.
\end{equation}
Standard computations transform \eqref{eq:constrained_variation} into the following condition
\begin{equation}
\label{eq:constrained_variation_sequel}
\begin{split}
	\int_{t_1}^{t_2}&
	\bigg\{\sum_{i=i}^{n-1}\bigg[\frac{\partial}{\partial q_i}\bigg(L-(\mu+\lambda)f\bigg)-\frac{d}{dt}\bigg(\frac{\partial}{\partial\dot{q}_i}\big(L-(\mu+\lambda)f\big) \bigg)\bigg]\delta q_i \\&
	-\frac{d}{dt}\bigg(\frac{\partial L}{\partial\dot{q}_n}+\lambda\bigg)\delta q_n\bigg\}dt=0.
\end{split}
\end{equation}
Since here all $\delta q_i$, for $i=1,\dots,n$\PPG{,} are independent, \eqref{eq:constrained_variation_sequel} requires that 
\begin{equation}\label{eq:lambda_constant}
\frac{\partial L}{\partial \dot q_n}+\lambda=\text{constant}.
\end{equation}
Evaluating \eqref{eq:lambda_constant} on \eqref{defcn}, which is the implicit version of the constraint \eqref{mono}, we conclude that also $\lambda$ is constant and so \eqref{eq:constrained_variation_sequel} reduces to the stationarity of the action $\mathcal{A}_{\widebar{L}}$ associated with a Lagrangian $\widebar{L}:=L-c_0\dot{q}_n$, where $c_0=\mu+\lambda$ is an arbitrary constant. It is a simple matter to check that setting $c_0=c_n$ has the noticeable advantage of making $\widebar{L}$ a function independent of $\dot{q}_n$ when evaluated on the constraint \eqref{mono}, as then
$$
\frac{\partial\widebar{L}}{\partial\dot{q}_n}=\frac{\partial L}{\partial \dot{q}_n}-c_n=0.
$$

Thus, if in the \emph{modified} Lagrangian
\begin{equation}\label{modla}
\widebar{L} :=L-c_n\dot q_n
\end{equation}
we make use of \eqref{mono}, we are guaranteed to obtain a function depending only on the non-cyclic coordinates (their derivatives and the parameter $c_n$), whose evolution is described by the usual Lagrange equations for a system with $n-1$ degrees of freedom. The complete evolution of the parent system with $n$ degrees of freedom is then obtained by quadrature of \eqref{mono}, once the evolution of the non-cyclic coordinates is known.
\begin{remark}\label{rmk:routh}
The \emph{modified} Lagrangian $\widebar{L}$ was first introduced by Routh in his essay \cite{routh:treatise} (see, in particular Sects.~20 and 21 of Chapt.~IV), where he also comments about his method being
%\begin{quote}
``equivalent to a partial use of Hamilton's transformation of Lagrange's equations.''
%\end{quote} 	
We also learn in Larmor's obituary of Routh \cite{larmor:obituary} that
\begin{quote}
	Lord Kelvin’s general theory of ``ignoration of co-ordinates,'' first published in 1879 in the second edition of Thomson and Tait’s treatise \cite{thomson:treatise} [\dots] probably existed in manuscript anterior to Routh’s essay. [\dots]
	%\end{quote}
	%However,
	%\begin{quote}
	This form of the theory, though more expressly suggested by the needs of physical dynamics, was less complete in one respect than Routh’s, in that it did not bring the matter into direct relation with a single characteristic function (Lagrangian function of Routh, kinetic potential of Helmholtz), but simply obtained and illustrated the equations of motion that arose from the elimination of the cyclic co-ordi­nates that could be thus ignored.
	Later still, Helmholtz, in his studies on monocyclic and polycyclic kinetic systems, which began in 1884 \cite{H1}, and culminated in the important memoir on the physical meaning of the Principle of Least Action in vol. c. (1886) of Crelle’s Journal \cite{helmholtz:physikalische}, developed the same theory more in Routh’s manner, and built round it an extensive discussion of physical phenomena, so that on the Continent the whole subject is usually coupled with his name. Shortly before, the work of Routh and Kelvin had already been co-ordinated with the Principle of Action by more than one writer in England.
\end{quote}
\end{remark}

\PPG{After having}  illuminated the connection between Routh's theory and the stationarity of the action, we apply Routh's method in a generic case.
%\[
%\overline L=\bar L(q_1,q_2,\ldots,q_{n-1}; \dot q_1, \dot q_2,\ldots, \dot q_{n-1};c_n;t),\quad L=L(q_1,q_2,\ldots,q_{n-1}; \dot q_1, \dot q_2,\ldots, \dot q_n;t),
%\]
%%
%and 
%where $\dot q_n$ is specified by (\ref{mono}).
We} split the Lagrangian (\ref{lagra3}) additively: 
\begin{equation}\label{eq:additive_splitting}
L=K-U, 
\end{equation}
with  
\begin{equation}\label{cinene}
K=\frac 1 2\sum_{i,k=1}^{n} a_{ik}\dot q_i\dot q_k\quad (a_{ik}=a_{ki})
\end{equation}
the \emph{kinetic energy}, as in \eqref{eq:kinetic_lego}, and 
\[
U=U(q_1,q_2,\ldots,q_{n-1};t)
\]
the \emph{potential energy}. Note that \eqref{cinene} can be written in the form:
\[
K=\frac 1 2\sum_{i,k=1}^{n-1} a_{ik}\dot q_i\dot q_k+\Big(\sum_{i=1}^{n-1} a_{in}\dot q_i\Big)\dot q_n+\frac 1 2 a_{nn}\dot q_n^2.
\]
Under the present circumstances, this further splitting implies that
\[
c_n=p_n=\frac{\partial K}{\partial \dot q_n}=\sum_{i=1}^{n-1} a_{in}\dot q_i+ a_{nn}\dot q_n\quad\Rightarrow\quad\dot q_n=a_{nn}^{-1}\Big(c_n-\sum_{i=1}^{n-1} a_{in}\dot q_i\Big).\footnote{The symmetric matrix whose entries we denoted by $a_{ik}$ is customarily taken positive definite, so that, in particular, all diagonal entries are positive.}
\]
Consequently,
\[
%\begin{aligned}
K-c_n\dot q_n=\frac 1 2\sum_{i,k=1}^{n-1} a_{ik}\dot q_i\dot q_k-\frac 1 2 a_{nn}^{-1}\Big(c_n-\sum_{i=1}^{n-1} a_{in}\dot q_i \Big)^2\\
\]
and, \EGV{by analogy with \eqref{eq:additive_splitting}}, we are driven to set:
\[
\widebar{L}=\widebar{K}-\widebar{U},
\]
with
\begin{align}
\widebar{K}&:=\frac 1 2\Bigg(\sum_{i,k=1}^{n-1} a_{ik}\dot q_i\dot q_k-a_{nn}^{-1}\Big(\sum_{i=1}^{n-1} a_{in}\dot q_i\Big)^2\Bigg)+a_{nn}^{-1}c_n\sum_{i=1}^{n-1} a_{in}\dot q_i ,\label{eq:K_bar}\\
\widebar{U}&:= U+\frac 1 2 a_{nn}^{-1}c_n^2.\label{eq:U_bar}
\end{align}
\EGV{
\begin{remark}
It is worth noting that a Lagrangian $L$ with a cyclic coordinate that contains only terms quadratic in the velocities is modified in a Lagrangian $\widebar{L}$ that also contains line\PPG{a}r terms. In Whittaker's terminology (see footnote~\ref{foot:wittaker}), we may say that  a cyclic coordinate drives a natural system into a non-natural one with one degree of freedom less.  
\end{remark}
}
\EGV{If the
kinetic coefficient} $a_{in}\neq 0$, the velocity $\dot{q}_n$ of the fast `ghost' coordinate and the velocity $\dot{q}_i$ of the slow coordinate  are \PPG{said to be} \emph{kinetically coupled}: a so-called \emph{gyroscopic term} $a_{in}\dot q_i$ is found in the kinetic energy $\widebar{K}$, a term which is linear in the rate $\dot q_i$ and does not have a definite sign (see \cite{L},  p.\,129). Furthermore, in the potential energy $\widebar{U}$ a positive contribution of kinetic origin is found, which morally accounts for the   variable $q_n$; in case of no kinetic coupling, this is the only manifestation of a microscopically fast variable, a candidate to be interpreted as \emph{stored heat content} in case $q_n$ is interpreted as thermal displacement.

\begin{remark}\label{rmk:thomson}
The connection between cyclic coordinates and hidden motions goes far beyond Helmholtz's works.
As lucidly explained by L\"utzen in his book \cite{lutzen:dissipative} (see, in particular, Chapter 18), cyclic coordinates have an interesting history that, \EGV{as already recalled in Remark~\ref{rmk:routh}, starts with Routh's method \cite{routh:treatise} and has far reaching consequences. In particular,}
\begin{comment}for deriving systematically reduced Lagrangian equations phrased into non-cyclic coordinates only.
\end{comment}
%, which in its essence has just been recalled,
this method was  used by J.J. Thompson in his papers \cite{thomson:some_1885,thomson:some_1887} and book \cite{thomson:applications} to interpret cyclic coordinates as manifestations of a hidden motion,  in accord with Helmholtz. Thompson considered a dynamical system whose energy was only kinetic and consisted of two parts, both quadratic, the one in non-cyclic and the other in cyclic coordinates only; he proved that the cyclic kinetic energy can be converted through Routh's method into an effective potential in the non-cyclic coordinates, thus showing that an ordinary potential energy can in principle be regarded as arising from a hidden motion. Liouville \cite{liouville:equations} went somehow the opposite way: he proved that any classical Lagrangian system, with kinetic and potential energies, can be converted into an equivalent one with only kinetic energy and a single added cyclic coordinate. 
\end{remark}	
\subsection{Hertzian  Mechanics} 
We read in \cite{L} (Chapter V, Sect. 4) that the fact that potential energy may include the kinetic contribution of a nonobservable  coordinate, \EGV{as shown for example in \eqref{eq:U_bar},} induced Hertz \cite{He} to dream of a \emph{forceless mechanics}; a cursory account of his views follows.

It helps to premise a classification of forces, in spite of the somewhat obsolete terminology. Within a Lagrangian/Hamiltonian framework there are two types of forces: 
\begin{enumerate}[(1)]
\item \emph{monogenic}, when their  \emph{incremental working}, \EGV{defined as} 
\[
dw^{(m)}=\sum_{i=1}^n F_i^{(m)} dq_i,
\]
is deducible from a scalar \emph{work function}  $V(q_1,\ldots,q_n;\dot q_1,\ldots,\dot q_n;t)$:
\[
dw^{(m)}=\sum_{i=1}^n \Big(\frac{\partial V}{\partial q_i}-\frac{d}{dt}\Big(\frac{\partial V}{\partial \dot q_i}\Big)\Big) dq_i\quad\Rightarrow\quad F_i^{(m)}:=\frac{\partial V}{\partial q_i}-\frac{d}{dt}\Big(\frac{\partial V}{\partial \dot q_i}\Big)\,;
\]
\PPG{The negative of the work function is identified with the potential energy: $U=-V$;  monogenic forces are said \emph{conservative} if their work function does not depend on $\dot q$ and $t$.}
\item \emph{polygenic}, otherwise, that is, when no work function can be associated to their incremental working
\begin{equation}\label{incwo}
dw^{(p)}=\sum_{i=1}^n F_i^{(p)} dq_i,
\end{equation}
\EGV{where the generalized forces $F_i^{(p)}$ are assumed to be functions of the  $q_i$'s and (possibly) of the $\dot{q}_i$'s too.}
\end{enumerate} 
%
%\sout{In both instances t}{\clr T}he negative of the work function is identified with the potential energy: $U=-V$;  monogenic forces are said \emph{conservative} if their work function does not depend on $\dot q$ and $t$.
%(NB. Qui ho invertito l'uso delle lettere $U$ e $V$ fatto da Lanczos). 

Monogenic forces are associated with  \emph{holonomic} kinematic conditions, in the sense that \PPG{contingent} kinematic conditions on a system's evolution are maintained by monogenic forces: on  p.\,114 of \cite{L}, we read: ``\ldots Hamilton's principle holds for arbitrary mechanical systems which are characterized by monogenic forces and holonomic auxiliary conditions''; the same assertion holds when a Lagrangian formulation is adopted. \emph{Nonholonomic} kinematic conditions call for polygenic forces; examples of the latter are inertia forces and friction forces. When a system is acted upon by noninertial polygenic forces, \EGV{in addition to monogenic forces whose work function $V$ is absorbed in the usual way in the Lagrangian $L$}, the \PPG{equations of motion are}
\begin{equation}\label{lagra2}
\frac{d\,}{dt}\left(\frac{\partial L}{\partial \dot q_i}\right)-\frac{\partial L}{\partial q_i}=F_i^{(p)} \quad(i=1,2,\ldots,n)
\end{equation}
(cf. equation (59.4), on  p.\,146 of \cite{L}).
\begin{comment}
As we shall see, classic M(olecular) D(ynamics) is Hamiltonian/Lagrangian; polygenic forces have citizenship in the more recent D(issipative)MD, where the equations of particle motion have the form \eqref{lagra2}.
\end{comment}
%(credo che questo sia il caso anche nelle simulazioni di Tildesley \& friends).

\PPG{In \cite{L},} Lanczos credits Hertz with the idea that a typical mechanical system  has many `hidden' ($\equiv$ nonobservable) degrees of freedom, that is, degrees of freedom that are not conveyed into the system's Lagrangian by a set of slow coordinates; and that there are two kinds of such hidden degrees of freedom, those associable with polygenic forces and those associable with  fast Lagrangian variables, whose presence is reflected in the kinetic portion of the potential energy and, possibly, in the gyroscopic terms of the kinetic energy. Apparently, according to Lanczos, Hertz went so far as to propound a forceless Hamiltonian mechanics, in which all of potential energy, just as kinetic energy itself, has a kinetic origin, in that it is induced by fast Lagrangian variables. 

% all of potential energy could be regarded as due to fast Lagrangian variables, in that   --che non mi pare condivisibile e che non credo sia stata condivisa granch\'e--\`e che tutta l'energia potenziale possa essere riguardata come dovuta a fast Lagrangian variables, cio\`e, che ogni forza monogenica non sia che la manifestazione in termini di energia potenziale di una qualche fast variable e, quindi, che l'energia potenziale abbia un'origine cinetica, proprio come l'energia cinetica.

\subsection{Helmholtz's  Heat Theorem}\label{hh2}
This theorem establishes a representation for the mechanical analogue of entropy; it can be seen as a tool to  explain \EGV{on the basis of the microscopic laws of mechanics the statement of the Second Law of the XIX century's thermodynamics of homogeneous systems to the effect  that \emph{coldness}, the inverse of temperature, is the integrating factor needed to reconstruct entropy from heat exchange  \cite{clausius:verschiedene,GG}}. We prove it in the simplest instance of a monocyclic Lagrangian system, whose coordinates are only two, one slow and one fast. 

Let the relevant Lagrangian be 
\[
L(q_s,\mathrlap{q_f}\bigtimes;\mathrlap{\dot q_s}\bigtimes,\dot q_f;t)=K(\dot q_f)-U(q_s)=\frac 1 2 p_f\dot q_f-U(q_s), \quad\textrm{with}\;\; p_f=m\,\dot q_f,
\]
and let the total incremental working be
\[
dw=g_f\,dq_f,
\]
\EGV{where $g_f$ appears to be a (generalized) force doing work against the fast variable $q_f$ and which we may assume to depend on $q_s$.}
Then, in view of \eqref{incwo} and \eqref{lagra2}, the motion equations are
\[
\frac{\partial L}{\partial q_s}=0,\quad \frac{d\,}{dt}\left(\frac{\partial L}{\partial \dot q_f}\right)-g_f=0\;\,\Leftrightarrow\;\,\dot p_f-g_f=0.
\]
%
%%
%\[
%\frac{\partial L}{\partial q_s}+g_s\!\!\!\!\!/\!\!\!\backslash\,=0,\quad -\frac{d\,}{dt}\left(\frac{\partial L}{\partial \dot q_f}\right)+g_f=0\;\,\Leftrightarrow\;\,-\dot p_f+g_f=0.
%\]
%%
Next, the incremental heat exchange is  equated to the incremental working performed on the system, and use is made of the motion equations:\cite{thomson:treatise}
\[
dQ=g_f dq_f=\dot p_f\dot q_f dt=\dot q_f dp_f;
\]
a division by $  p_f\dot q_f=2K$ yields
\[
\frac{dQ}{2\,K}=d(\log p_f).
\]
With this, on setting
\begin{equation}\label{eq:temperature_lego}
2\,K=:\text{temperature} \;\,\temp\quad \textrm{and}\quad \log p_f:=\text{entropy} \;\,S,
\end{equation}
we have the desired conclusion:
\[
\frac{dQ}{\temp}=dS,
\]
with $\temp^{-1}$, the `coldness', as the integrating factor. Note that this result does not depend on the form of the potential energy.

\section{Analytical Thermodynamics of Homogeneous Systems}\label{sec:homo}
Here, building upon Helmholtz's mechanistic interpretation, we introduce a formal analytic theory for the thermodynamics of \emph{homogeneous} systems. \EGV{Our aim is to provide a theoretical setting for \emph{nonequilibrium} thermodynamics, within which the evolution out of equilibrium can be derived from the Lagrangian/Hamiltonian formalism illustrated above. Our approach will be constructive: starting from the basic potentials of classical thermodynamics, namely, free energy, entropy, and dissipation (or entropy production), we seek  Lagrangians (and Hamiltonians) that govern evolution out of equilibrium consistently with the law of thermodynamics.}\footnote{\EGV{This approach characterizes the meaning that we properly attach to \emph{analytical thermodynamics}; it differs from that implied by Li in the title of his book \cite{li:analytical}, where \emph{analytical} is meant to evoke the solid, theoretical structure provided to thermodynamics since the pioneering, elegant work of Gibbs \cite{gibbs:elementary} (see p.\,v of \cite{li:analytical}).}}

\EGV{Homogeneous} systems \EGV{have} admissible thermodynamic states \EGV{that} can be described by $n-1$ variables, changing with time but uniform in space, the entries of a state vector $\PPG{ q_s=(q_1,\dots,q_{n-1})}\in\Omega$, where $\Omega\subset\mathbb{R}^{\PPG{ n-1}}$ is a given admissible set. In the language introduced in Sect. \ref{HMIT}, \PPG{ the vector $q_s$ consists of \emph{slow} variables, whose}
%these variables are \emph{slow}; their 
time derivatives do not feature in the Lagrangian. We assume that a single \emph{fast} variable, $\PPG{ q_f}$, of a microscopic origin and unobservable at the macroscopic scale, produces observable macroscopic effects through its time derivative $\PPG{ \dot{q}_f}$, which we phantom related to the temperature $\vartheta$ of the system, in a manner  that will be made precise. Thus, the Lagrangian $L$ will be a function 
$L=L(\PPG{q_s};\cycl)$.
%\begin{remark}
%Here our notation slightly departs from that in Sect.~\ref{HMIT}, as \PPG{ now} the state vector 
%%$\PPG{ q_s}$ 
%only comprises slow variables \PPG{ and there is only one fast variable}. {\clb Forse questo rmk si potrebbe sopprimere.}
%%To reconcile the two notations, the reader may set $m=n-1$ and identify $q_n$ with $\PPG{ q_s}$ (and, consequently, $\dot{q}_n$ with $\dot{q}_0$).
%\end{remark}	

In Helmholtz's terminology, the typical homogeneous system we \EGV{envision} is \emph{monocyclic}, governed by a Lagrangian action principle which we extend both in a d'Alembert fashion---so as to incorporate nonconservative thermodynamic
forces, depending only on the state vector $\PPG{ q_s}$ and collected in a vector $\PPG{ Q_s}=(Q_1,\dots,Q_{n-1})$---and in a Rayleigh fashion---so as to incorporate also dissipative forces. 
depending on $\PPG{ q_s}$ and $\PPG{ \dot q_f}$ and 
derived from a potential $R=R(\PPG{ q_s};\PPG{\dot{q}_s},\cycl)$ which depends also on the rate of the fast coordinate $\PPG{q_f}$.
Precisely,
% on l}etting the non-conservative positional forces have Lagrangian components $Q_i$, so as to form the vector $Q=(Q_1,\dots,Q_n)$, 
we require the stationarity of action $\mathcal{A}_L$ augmented by the virtual work associated with all forces and the fast coordinate:
\begin{equation}
\label{eq:Hamilton_principle_augmented}
\delta\mathcal{A}_L+\int_{t_0}^{t_1}\left(\PPG{ Q_s}\cdot\delta \PPG{ q_s}-
\frac{\partial R}{\partial\PPG{\dot{q}_s}}\EGV{\cdot} \delta\PPG{ q_s}-\frac{\partial R}{\partial\cycl}\delta \PPG{ q_f}\right)dt=0.
\end{equation}
\begin{remark} %in the dissipative Lagrangian density, 
We regard the generalized velocities $(\dot{q}_s,\dot{q}_f)$ as %treated as 
\emph{neutral} to the variation indicated in \eqref{eq:Hamilton_principle_augmented}
(as effectively said in \cite[p.\,167]{gyarmati:nonequilibrium}) and hence we keep them constant during that variation (in tune with the way \emph{variational} constraints are treated in \cite{gay-balmaz:lagrangian_review}).
\end{remark}
By the special form of $L$, \PPG{in} \eqref{eq:Hamilton_principle_augmented} 
\begin{equation}
\label{eq:eq:Hamilton_principle_special}
\delta\mathcal{A}_L=\int_{t_0}^{t_1}\left\{-\left(\frac{d}{dt}\frac{\partial L}{\partial\cycl}\right)\delta \PPG{ q_f}+\frac{\partial L}{\partial \PPG{ q_s}}\cdot\delta \PPG{ q_s}\right\}dt;
\end{equation}
in accord with our discussion in Sect.~\ref{HMIT}, $\delta \PPG{ q_f}$ is a kind of \emph{virtual} thermal displacement. Moreover, $\PPG{ Q_s}$ has a purely mechanical origin \EGV{and we think of it as a function of $q_s$ only}: there is no force performing virtual work against a thermal displacement.

The evolution (dynamical) equations associated with the variational principle stated in \eqref{eq:Hamilton_principle_augmented} are obtained by requiring that stationarity is achieved there for arbitrary variations $(\delta \PPG{ q_s},\delta \PPG{ q_f})$; these equations read as
\begin{subequations}\label{eq:dynamical_equations}
\begin{align}
\frac{d}{dt}\frac{\partial L}{\partial\cycl}+\frac{\partial R}{\partial\cycl}&=0,\label{eq:dynamical_equation_1}\\
\frac{\partial L}{\partial\PPG{q_s}}-\frac{\partial R}{\partial\PPG{\dot q_s}}+\PPG{Q_s}&=0.
\label{eq:dynamical_equation_2}
\end{align}
\end{subequations}

In \eqref{eq:Hamilton_principle_augmented}, as in \cite{virga:rayleigh} and unlike what is customary, we do \EGV{\emph{not}} assume that \EGV{$R(q_s;\cdot,\cdot)$} is a quadratic form \EGV{for all $q_s$};\footnote{\EGV{This assumption would guarantee that \eqref{eq:dynamical_equation_2} could be reduced to normal form, which is \emph{not} explicitly contemplated in our theory.}} we want only to interpret it as a \emph{dissipation} potential, characterized by the property
\begin{equation}
\label{eq:dissipation_potential}
\PPG{\dot{q}_s}\cdot\frac{\partial R}{\partial\PPG{\dot{q}_s}}+\cycl\frac{\partial R}{\partial\cycl}=\dis(\PPG{q_s};\PPG{\dot{q}_s},\cycl),
\end{equation}
where $\dis$, the dissipation, is an assigned function, positive semidefinite in  $(\PPG{\dot{q}_s},\cycl)$.\footnote{\EGV{As customary, by this we simply mean that $\dis(q_s;\cdot,\cdot)\geqq0$ for all choices of  $q_s$.}} As pointed out in \cite{virga:rayleigh}, the Rayleigh potential $R$ can then be easily retrieved through the formula
\begin{equation}
\label{eq:dissipation_formula}
R(\EGV{q_s};\PPG{\dot{q}_s},\cycl)=\left.\int\dis\left(\PPG{ q_s};e^s\PPG{ \dot{q}_s},e^s\cycl\right)ds\right|_{s=0};
\end{equation}
whenever $\dis$ is a homogeneous function of degree $d$ in $(\PPG{\dot{q}_s},\cycl)$ (an assumption which is not needed here), relation \eqref{eq:dissipation_formula}  delivers
\begin{equation}
\label{eq:dissipation_special}
R=\frac{1}{d}\dis\,.
\end{equation}

The role of $\dis$ is fully appreciated within a thermodynamic framework, such as the one expounded by Truesdell \cite[pp.\,9-10]{truesdell:rational} (see also \cite[p.\,126]{sonnet:dissipative}). In one of the many ways to formulate the \emph{Second Law} of thermodynamics, we may say that $\dis$ \emph{fills the gap} between the \emph{heating} $\heat$ provided to the system and the rate of entropy growth $\dot{\ent}$, as
\begin{equation}
\label{eq:second_law}
\vartheta\dot{\ent}-\heat=\dis\geqq0,
\end{equation}
where $\vartheta$ is the absolute temperature.\footnote{In Truesdell's words \cite[p.\,9]{truesdell:rational}, ``We assume the existence of a second kind of working, $\heat$, called \emph{heating}, which is not identified with anything from mechanics'' (see also \cite[p.\,125]{sonnet:dissipative}).}
The \emph{First Law} of thermodynamics is %accordingly 
formulated as 
%a {\color{red}split balance} of the different forms of energy:
\begin{equation}\label{eq:first_law}
\dot\ene=\work+\heat,%\quad \work=\work^{in}+\work^{ni},
\end{equation}	
where the rate of  \emph{internal energy} $\ene$ is balanced by the mechanical power expended by all nondissipative agencies 
$\work$ and the thermal power $\heat$.
%\footnote{ Under the standard  assumption that
%%
%\[
%{\color{red}\dot\kin+\work^{in}=0},
%\]
%%
%\eqref{eq:first_law} can be written as
%%
%\[
%{\color{red}\dot\ene}=\work+\heat,\quad \work=\work^{in}+\work^{ni},
%\end{equation}where $\kin$ kinetic energy, and $\work^{in}$ power expended by  {\color{red}inertial} agencies {\color{red}qui possiamo citare \emph{Inertia and Invariance} a proposito della seconda \eqref{eq:first_law}. }}
Letting 
\begin{equation}
\label{eq:free_energy}
\free:=\ene-\vartheta\ent
\end{equation}
denote Helmholtz's \emph{free energy}, and making use of \eqref{eq:first_law}, we rewrite \eqref{eq:second_law} in the form
\begin{equation}
\label{eq:second_law_rewritten}
\work-\dot{\free}-\dot\vartheta\ent=\dis\geqq0.
\end{equation}
%{\color{red} qui si potrebbe citare il metodo di Coleman and Noll per decidere l'ammissibilit\`a termodinamica, sottolineando che non \`e necessario decidere la forma di $\dis$}

To lay the basis of our analytical thermodynamics governed by \eqref{eq:dynamical_equations}, we need to establish a relation with classical thermodynamics that goes beyond the natural identification in \eqref{eq:dissipation_formula}. We need to link $L$ to $\free$ and $\ent$, for which we assume the following constitutive relations
\begin{eqnarray}
\label{eq:constitutive_relations}
\free=\free(\vartheta;\PPG{ q_s})\quad\text{and}\quad\ent=\ent(\vartheta;\PPG{ q_s}),
\end{eqnarray}
which are in tune with the classical view of considering both $\free$ and $\ent$ as functions depending on temperature and thermodynamic state. We also need to link $\cycl$ to $\vartheta$; for the time being, we assume that
\begin{equation}
\label{eq:f_definition}
\vartheta=f(\cycl),
\end{equation}
as most of the theory developed below will be independent of the specific choice of the function $f$.

We readily obtain \EGV{from \eqref{eq:dynamical_equations} and \eqref{eq:dissipation_potential}} that
\begin{equation}
\label{eq:power_balance}
-\frac{d}{dt}\left(\frac{\partial L}{\partial\cycl}\right)\cycl=\dis-\frac{\partial L}{\partial \PPG{ q_s}}\cdot \PPG{\dot{q}_s}-\PPG{ Q_s}\cdot\PPG{\dot{q}_s}=-\dot{\free}-\dot{\vartheta}\ent,
\end{equation}
where we have made use of \eqref{eq:second_law_rewritten} and identified $\work$ as the working of all nondissipative thermodynamic forces (both conservative and non-conservative),
\begin{equation}
\label{eq:working_identification}
\work=\left(\frac{\partial L}{\partial \PPG{ q_s}}+\PPG{ Q_s}\right)\cdot\PPG{ \dot{q}_s}.
\end{equation}
By expanding the first and last terms in \eqref{eq:power_balance}, with the aid of \eqref{eq:constitutive_relations} and \eqref{eq:f_definition}, we arrive at
\begin{equation}
\label{eq:identity}
\frac{\partial^2L}{\partial\cycl^2}\PPG{\dot{q}_f}\PPG{\ddot{q}_f}+\frac{\partial}{\partial\PPG{ q_s}}\left(\cycl\frac{\partial L}{\partial\cycl}\right)\cdot\PPG{ \dot{q}_s}=\left(\frac{\partial\free}{\partial\vartheta}+\ent\right)f'(\cycl)\PPG{\ddot{q}_f}+\frac{\partial\free}{\partial \PPG{ q_s}}\cdot\PPG{\dot{q}_s}.
\end{equation}
Requiring the linear forms in $(\EGV{\ddot{q}_f,\dot{q}_s})$ on the two sides of this equation to be identical, we derive the equations
\begin{subequations}
\label{eq:identities}
\begin{align}
\left(\frac{\partial\free}{\partial\vartheta}+\ent\right)f'&=\frac{\partial^2L}{\partial\cycl^2}\cycl,\label{eq:identity_1}\\
\free(f(\cycl);\EGV{q_s})&=\cycl\frac{\partial L}{\partial\cycl}+G(\cycl),\label{eq:identity_2}
\end{align}
\end{subequations}
where $G$ is an arbitrary function. Differentiating both sides of \eqref{eq:identity_2} with respect to $\EGV{\dot{q}_f}$ and using \eqref{eq:identity_1}, we conclude that
\begin{equation}
\label{eq:G_constraint}
\ent f'=-\frac{\partial}{\partial\cycl}(L+G),
\end{equation}
whence it follows that
\begin{equation}
\label{eq:G_formula}
L(\EGV{q_s};\cycl)=-\int_0^{f(\cycl)}\ent(\tau;\PPG{ q_s})d\tau\EGV{-G(\cycl)}-A(\PPG{ q_s}),
\end{equation}
where $A$ is an arbitrary function. 
\EGV{To determine $G$, we make use of \eqref{eq:G_formula} in  \eqref{eq:identity_1}, obtaining that
\begin{equation}
\label{eq:G_equation}
\dot{q}_f\frac{\partial^2G}{\partial\dot{q}_f^2}=-\left(\frac{\partial\free}{\partial\temp}+\ent\right)f'-\left(\frac{\partial\free}{\partial\temp}f'^2+\ent f''\right)\dot{q}_f,
\end{equation}
which can be solved by quadrature for $G$, provided that the following compatibility condition is obeyed\PPG{:}
\begin{equation}
\label{eq:compatibility_condition}
\frac{\partial}{\partial q_s}\left\{f'\left(\frac{\partial\free}{\partial\temp}+\ent\right)+f'^2\dot{q}_f\frac{\partial\ent}{\partial\temp}+f''\dot{q}_f\ent \right\}=0.
\end{equation}
\begin{remark}\label{rmk:compatibility}
Condition \eqref{eq:compatibility_condition} is rather intricate as it also involves the unknown function $f$ expressing $\temp$ as in \eqref{eq:f_definition}. We can give it a more transparent expression by using the inverse function $\phi$ of $f$, defined by
\begin{equation}
	\label{eq:phi_definition}
	\dot{q}_f=\phi(\temp).
\end{equation}
The following identities are immediate consequences of \eqref{eq:f_definition} and \eqref{eq:phi_definition},
$$
\phi'f'=1\quad\text{and}\quad \phi''f'^2+\phi'f''=0,
$$
where a prime $'$ denotes differentiation.
Simple computations then show that \eqref{eq:compatibility_condition} is equivalent to
\begin{equation}
	\label{eq:compatibility_condition_reduced}
	\frac{\phi'}{\phi}\left(\frac{\partial\free}{\partial\temp}+\ent\right)+\frac{\partial\ent}{\partial\temp}-\frac{\phi''}{\phi'}\ent=\Sigma(\temp),
\end{equation}
with $\Sigma$ an arbitrary function. Given $\free$ and $\ent$, \eqref{eq:compatibility_condition_reduced} can be interpreted as an equation for $\phi$.
\end{remark}
\begin{remark}
\label{rmk:ideal_gas}
It might be a bit disheartening to realize that there is no function $\phi$ satisfying \eqref{eq:compatibility_condition_reduced} for an ideal polyatomic gas. Indeed,  by taking (to within inessential additive constants)
\begin{equation}
	\label{eq:ideal_gas}
	\ent=\frac{M}{m}k\ln\left(\temp^3V\right)\quad\text{and}\quad\ene=\frac{M}{m}3k\temp,
\end{equation}
where $k$ is the Boltzmann constant, $M$ is the total mass of the gas, $m$ the mass of a gas molecule, and $V$ the volume occupied by the system (see, for example, p.\,9 of \cite{mueller:entropy}), we easily see that \eqref{eq:compatibility_condition_reduced} is violated for any $\phi$, \emph{if} $V$ is to be interpreted as a $q_s$ variable.
We have no clear justification for such a disappointing result.
\end{remark}
}

Combining equation \eqref{eq:G_formula} with \eqref{eq:identity_2} to eliminate $G$, and carrying out under the present circumstances the general developments in the first part of Sect. \ref{LvsH}, we find  the Hamiltonian $H(\EGV{p_f};\PPG{ q_s})$ conjugated with $L$ in terms  of $\free$ and $\ent$,
\begin{equation}
\label{eq:Hamiltonian}
H(\EGV{p_f};\PPG{ q_s})=\free(\vartheta;\PPG{ q_s})+\int_0^\vartheta\ent(\tau;\PPG{ q_s})d\tau+A(\PPG{ q_s}),
\end{equation}
where $\vartheta$ is now meant to be a positive function $\EGV{\widetilde{f}}$  of the kinetic momentum $p_f$ conjugated with  $\EGV{\dot{q}_f}$ \EGV{and possibly of  $q_s$.
\begin{remark}\label{rkm:p_f}
According to \eqref{mom}, it follows from \eqref{eq:G_formula} that
\begin{equation}
	\label{eq:p_f}
	p_f=-\frac{\partial G}{\partial \dot{q}_f}-\ent(f(\dot{q}_f);q_s)f'(\dot{q}_f).
\end{equation}
Assuming that \eqref{eq:p_f} can be inverted to deliver $\dot{q}_f$, we see from \eqref{eq:f_definition} that $\widetilde{f}$ does in general depend on both  $p_f$ and $q_s$.
\end{remark}
}

\EGV{Our task is thus accomplished: we have found in \eqref{eq:G_formula} a whole class of Lagrangians whose dynamical equations \eqref{eq:dynamical_equations} represent evolutions out of equilibrium consistent with the basic laws of classical thermodynamics. It is perhaps remarkable that every compatible Lagrangian $L$ depends directly on the entropy $\ent$ and indirectly (via $G$ in \eqref{eq:G_equation}) on the free energy $\free$, while the associated Hamiltonian $H$ has a more transparent expression \eqref{eq:Hamiltonian} featuring both  $\ent$ and $\free$.}
\begin{remark}
In equilibrium thermodynamics, $\ent$ and $\free$ are related through the equation
\begin{equation}
\label{eq:entropy_equilibrium}
\ent=-\frac{\partial\free}{\partial\vartheta}
\end{equation}
\PPG{(}see, for example, \cite[p.\,127]{swendsen:introduction}\PPG{ )}. Now, in equilibrium all dissipative forces cease to act and Rayleigh-Lagrange dynamics is not different from Hamiltonian dynamics. Thus, by applying \eqref{hamit} with $H$ as in
\eqref{eq:Hamiltonian}, we readily recover \eqref{eq:entropy_equilibrium} and  add a further  requirement,
\begin{equation}
\label{eq:equilibrium_q}
\frac{\partial\free}{\partial \PPG{ q_s}}+\frac{\partial}{\partial \PPG{ q_s}}\int_0^\vartheta\ent(\tau;\PPG{ q_s})d\tau+\frac{\partial A}{\partial \PPG{ q_s}}=0,
\end{equation}
that is, an equation for the equilibrium state vector.
\end{remark}
\begin{remark}
It is perhaps worth noting that, were \eqref{eq:entropy_equilibrium} adopted as a constitutive law for $\ent$, \eqref{eq:Hamiltonian} would deliver a Hamiltonian $H_0$ independent  of $\EGV{p_f}$,
\begin{equation}
\label{eq:Hamiltonian_zero}
H_0(\PPG{ q_s})=\free(0;\PPG{ q_s})+A(\PPG{ q_s}).
\end{equation}
In such an instance, equation \eqref{eq:equilibrium_q} reduces to the customary requirement of stationarity for the free energy, if we set $A\equiv\text{constant}$ and we regard it \EGV{as} a \emph{gauge} function for the free energy.
\end{remark}
\begin{remark}	
As a simple application of our method, we now recover the classical case considered by Onsager and Machlup \cite{onsager:fluctuations} for the evolution of irreversible processes near equilibrium. We assume that \eqref{eq:entropy_equilibrium} be satisfied near equilibrium as well, so that the Hamiltonian can be written as a function of the state vector $\PPG{ q_s}$ only, as in \eqref{eq:Hamiltonian_zero}. Specifically, we take
\begin{equation}
\label{eq:H_0_Onsager}
H_0=\frac12\sum_{i,j=1}^{n-1}h_{ij}q_iq_j,
\end{equation}
where $h_{ij}$ is a symmetric, positive definite matrix. If, as in \cite{onsager:fluctuations}, we write the Rayleigh potential as 
\begin{equation}
\label{eq:R_Onsager}
R=\frac12\sum_{i,j=1}^{n-1}r_{ij}\dot{q}_i\dot{q}_j,
\end{equation}
where $r_{ij}$ is another symmetric, positive definite matrix, independent of $\PPG{ q_s}$, we readily see that equation \eqref{eq:dynamical_equation_1} is identically satisfied, while equation \eqref{eq:dynamical_equation_2} become
\begin{equation}
\label{eq:evolution_Onsager}
\sum_{j=1}^{n-1}(r_{ij}\dot{q}_j+h_{ij}q_j)=0,\quad i=1,\dots,n-1,
\end{equation}
which have the same form as the phenomenological laws (2-13) of \cite{onsager:fluctuations}, the only difference being that the matrix $h_{ij}$ is there replaced by the matrix $s_{ij}$ representing the quadratic approximation of the entropy near equilibrium (see also \cite{serdyukov:extension}).
\end{remark}

\section{Conclusion}\label{sec:conclusion}
We have proposed a thermodynamic theory that builds on the formalism of classical analytical mechanics.

We have adopted Helmholtz's mechanical interpretation of thermodynamics with its typical distinction between slow and fast variables, the latter characterized as thermal. Truesdell's thermodynamic theory, with its classical interpretation of entropy, has set the background, the connection with analytical mechanics being provided by Rayleigh's dissipation potential, here extended beyond its native quadratic limitation.

Our major result was to derive a whole class of Hamiltonians (each defined to within an arbitrary state function),  such that the associated evolution equations could be established in accord  with constitutive laws for both free energy and entropy, which in our nonequilibrium setting are unrelated.
As a first  application of our method, we showed that near equilibrium it delivers Onsager's linear laws for irreversible processes.

\EGV{At least two issues remain unsettled. First, a certain degree of indeterminacy for both Lagrangian and Hamiltonian associated with a given constitutive choice of free energy and entropy. How is the evolution away from equilibrium described by \eqref{eq:dynamical_equations} affected by this?

Second, apart from a cursory incursion into Onsager and Machlup's theory, we have not yet applied our proposed method to the evolution of some simple system, as are, for example,  those considered  in \cite{cendra:elementary}. This should be done, although Remark~\ref{rmk:ideal_gas} would suggest that such a task might require more ingenuity than expected.

More ambitiously,   
we also plan to apply our theory to  cases relevant to  the promising field of soft matter systems, where Onsager's principle has recently seen a surge of renewed interest. A reliable guidance to these problems  can be found in the recent review  \cite{wang:generalized}.}

\EGV{Finally, even a more futuristic avenue  for further exploration could be forseen}.  We treated only homogeneous systems, with a finite number of state variables depending only on time. An extension of the theory to continuous fields, which is presently lacking, would be desirable.

\bibliographystyle{spmpsci}      % mathematics and physical sciences
%\bibliography{AT}

\begin{thebibliography}{10}
\providecommand{\url}[1]{{#1}}
\providecommand{\urlprefix}{URL }
\expandafter\ifx\csname urlstyle\endcsname\relax
\providecommand{\doi}[1]{DOI~\discretionary{}{}{}#1}\else
\providecommand{\doi}{DOI~\discretionary{}{}{}\begingroup
\urlstyle{rm}\Url}\fi

\bibitem{arnold:mathematical}
Arnold, V.I., Neishtadt, A.I., Kozlov, V.V.: Mathematical Aspects of Classical
and Celestial Mechanics, \emph{Encyclopaedia of Mathematical Sciences},
vol.~3, 3 edn.
\newblock Springer, Berlin (2006).
\newblock Original {R}ussian edition (2nd ed.) published by URSS, Moscow 2002

\bibitem{baldiotti:hamiltonian}
Baldiotti, M.C., Fresneda, R., Molina, C.: A {H}amiltonian approach to
thermodynamics.
\newblock Ann. Phys. \textbf{373}, 245--256 (2016).
\newblock \doi{https://doi.org/10.1016/j.aop.2016.07.004}

\bibitem{biot:variational}
Biot, M.A.: Variational principles in irreversible thermodynamics with
application to viscoelasticity.
\newblock Phys. Rev. \textbf{97}(6), 1463--1469 (1955)

\bibitem{biot:variational_book}
Biot, M.A.: Variational Principles in Heat Transfer: Unified {L}agrangian
Analysis of Dissipative Phenomena.
\newblock Oxford University Press, Oxford (1970)

\bibitem{biot:virtual}
Biot, M.A.: A virtual dissipation principle and {L}agrangian equations in
non-linear irreversible thermodynamics.
\newblock Bulletin de l'Acad\'emie royale de Belgique (Classe des Sciences)
\textbf{61}, 6--30 (1975).
\newblock Also available at
http://www.pmi.ou.edu/Biot2005/biotConferenceBiotsPapers.htm

\bibitem{Bro2}
de~Broglie, L.: Sur la th\'eorie des foyers cin\'etiques dans la
thermodynamique de la particule isol\'ee.
\newblock Compt. R. Acad. Sci. Paris \textbf{257}, 1822--1824 (1963)

\bibitem{Bro1}
de~Broglie, L.: Sur l'introduction de l'\'energie libre dans la thermodynamique
cach\'ee des particules.
\newblock Compt. R. Acad. Sci. Paris \textbf{257}, 1430--1433 (1963)

\bibitem{Bro3}
de~Broglie, L.: La Th\'ermodynamique de la Particule Isol\'ee.
\newblock Gauthier-Villars, Paris (1964)

\bibitem{cendra:elementary}
Cendra, H., Grillo, S., {Palacios Amaya}, M.: Elementary thermo-mechanical
systems and higher order constraints.
\newblock Qual. Theory Dyn. Syst. \textbf{39}, 19 (2020).
\newblock \doi{https://doi.org/10.1007/s12346-020-00371-8}

\bibitem{clausius:verschiedene}
Clausius, R.: {\"U}ber verschiedene f\"ur die {A}nwendung bequeme {F}ormen der
{H}auptgleichungen der mechanischen {W}\"armetheorie.
\newblock Poggendorff's Ann. Physik \textbf{125} (1864)

\bibitem{coquinot:general}
Coquinot, B., Morrison, P.J.: A general metriplectic framework with application
to dissipative extended magnetohydrodynamics.
\newblock J. Plasma Phys. \textbf{86}(3), 835860302 (2020).
\newblock \doi{10.1017/S0022377820000392}

\bibitem{dirac:generalized}
Dirac, P.A.M.: Generalized hamiltonian dynamics.
\newblock Canadian J. Math. \textbf{2}, 129--148 (1950).
\newblock \doi{10.4153/CJM-1950-012-1}

\bibitem{doi:variational}
Doi, M.: Variational principle for the {K}irkwood theory for the dynamics of
polymer solutions and suspensions.
\newblock J. Chem. Phys. \textbf{79}, 5080--5087 (1983).
\newblock \doi{10.1063/1.445604}

\bibitem{doi:onsager}
Doi, M.: Onsager's variational principle in soft matter.
\newblock J. Phys.: Condens. Matter \textbf{23}, 284118 (2011).
\newblock \doi{10.1088/0953-8984/23/28/284118}

\bibitem{doi:onsager_tool}
Doi, M.: Onsager principle as a tool for approximation.
\newblock Chinese Physics B \textbf{24}, 020505 (2015).
\newblock \doi{10.1088/1674-1056/24/2/020505}

\bibitem{doi:onsager_polymer}
Doi, M.: Onsager principle in polymer dynamics.
\newblock Progr. Polym. Sci. \textbf{112}, 101339 (2021).
\newblock \doi{https://doi.org/10.1016/j.progpolymsci.2020.101339}

\bibitem{doi:application}
Doi, M., Zhou, J., Di, Y., Xu, X.: Application of the {O}nsager-{M}achlup
integral in solving dynamic equations in nonequilibrium systems.
\newblock Phys. Rev. E \textbf{99}, 063303 (2019).
\newblock \doi{10.1103/PhysRevE.99.063303}

\bibitem{eldred:single}
Eldred, C., Gay-Balmaz, F.: Single and double generator bracket formulations of
multicomponent fluids with irreversible processes.
\newblock J. Phys. A: Math. Theor. \textbf{53}(39), 395701 (2020).
\newblock \doi{10.1088/1751-8121/ab91d3}

\bibitem{GG}
Gallavotti, G.: Statistical Mechanics: a Short Treatise.
\newblock Springer, Berlin (1999)

\bibitem{gallavotti:elements}
Gallavotti, G.: The Elements of Mechanics, 2 edn.
\newblock Ipparco, Rome (2007).
\newblock The first edition was published by Springer-Verlag in 1983.

\bibitem{Galley}
Galley, C.R.: Classical mechanics of nonconservative systems.
\newblock Phys. Rev. Lett. \textbf{110}, 174301 (2013).
\newblock \doi{10.1103/PhysRevLett.110.174301}

\bibitem{gambar:hamilton}
Gamb\'ar, K., M\'arkus, F.: Hamilton-{L}agrange formalism of nonequilibrium
thermodynamics.
\newblock Phys. Rev. E \textbf{50}, 1227--1231 (1994).
\newblock \doi{10.1103/PhysRevE.50.1227}

\bibitem{gay-balmaz:lagrangian_I}
Gay-Balmaz, F., Yoshimura, H.: A {L}agrangian variational formulation for
nonequilibrium thermodynamics. {P}art {I}: {D}iscrete systems.
\newblock J. Geom. Phys. \textbf{111}, 169--193 (2017).
\newblock \doi{https://doi.org/10.1016/j.geomphys.2016.08.018}

\bibitem{gay-balmaz:lagrangian_II}
Gay-Balmaz, F., Yoshimura, H.: A {L}agrangian variational formulation for
nonequilibrium thermodynamics. {P}art {II}: {C}ontinuum systems.
\newblock J. Geom. Phys. \textbf{111}, 194--212 (2017).
\newblock \doi{https://doi.org/10.1016/j.geomphys.2016.08.019}

\bibitem{gay-balmaz:variational}
Gay-Balmaz, F., Yoshimura, H.: A variational formulation of nonequilibrium
thermodynamics for discrete open systems with mass and heat transfer.
\newblock Entropy \textbf{20} (2018).
\newblock \doi{10.3390/e20030163}

\bibitem{gay-balmaz:lagrangian_review}
Gay-Balmaz, F., Yoshimura, H.: From {L}agrangian mechanics to nonequilibrium
thermodynamics: {A} variational perspective.
\newblock Entropy \textbf{21} (2019).
\newblock \doi{10.3390/e21010008}

\bibitem{gibbs:elementary}
Gibbs, J.W.: Elementary Principles in Statistical Mechanics.
\newblock Charles Scribners's Sons, New York (1902).
\newblock Digitally reprinted by Cambridge University Press, Cambridge, in
2010.

\bibitem{green:re-examination}
Green, A.E., Naghdi, P.M.: A re-examination of the basic postulates of
thermomechanics.
\newblock Proc. R. Soc. Lond. A \textbf{432}, 171--194 (1991).
\newblock \doi{10.1098/rspa.1991.0012}

\bibitem{grmela:bracket}
Grmela, M.: Bracket formulation of dissipative fluid mechanics equations.
\newblock Phys. Lett. A \textbf{102}(8), 355--358 (1984).
\newblock \doi{https://doi.org/10.1016/0375-9601(84)90297-4}

\bibitem{grmela:dynamics}
Grmela, M., \"Ottinger, H.C.: Dynamics and thermodynamics of complex fluids.
{I}. {D}evelopment of a general formalism.
\newblock Phys. Rev. E \textbf{56}, 6620--6632 (1997).
\newblock \doi{10.1103/PhysRevE.56.6620}

\bibitem{gurtin:general}
Gurtin, M.E., Pipkin, A.C.: A general theory of heat conduction with finite
wave speeds.
\newblock Arch. Rational Mech. Anal. \textbf{31}, 113--126 (1968).
\newblock \doi{https://doi.org/10.1007/BF00281373}

\bibitem{gyarmati:nonequilibrium}
Gyarmati, I.: Non-equilibrium Thermodynamics: {F}ield Theory and Variational
Principles.
\newblock Springer-Verlag, New York (1970).
\newblock Originally published in Hungarian in 1967; translated into English by
E. Gyarmati and W. F. Heinz.

\bibitem{H1}
von Helmholtz, H.: Prinzipien der statik monocyklischer systeme.
\newblock Borchardt-Crelle's Journal f\"ur die reine und angewandte Mathematik
\textbf{97}, 111--140 (1884).
\newblock Also in Wiedemann G. (Ed.) (1895) Wissenschafltliche Abhandlungen.
Vol. 3 (pp. 142-162, 179-202). Leipzig: Johann Ambrosious Barth.

\bibitem{H2}
von Helmholtz, H.: Studien zur statik monocyklischer systeme.
\newblock Sitzungsberichte der K\"oniglich Preussischen Akademie der
Wissenschaften zu Berlin \textbf{I}, 159--177 (1884).
\newblock Also in Wiedemann G. (Ed.) (1895) Wissenschafltliche Abhandlungen.
Vol. 3 (pp. 163-178). Leipzig: Johann Ambrosious Barth.

\bibitem{helmholtz:physikalische}
von Helmholtz, H.: \"uber die physikalische {B}edeutung des {P}rincips der
kleinsten {W}irkung.
\newblock Journal f\"ur die reine und angewandte {M}athematik. {J}ournal de
{C}relle. {B}erlin. \textbf{100}, 213--222 (1886)

\bibitem{He}
Hertz, H.: Die Principie der Mechanik in neuem Zusammenhange dargestellt.
\newblock Barth, Leipzig (1894).
\newblock English translation: The Principles of Mechanics Presented in a New
Form, Macmillan, 1900. Reprinted Dover, New York 1950.

\bibitem{ichiyanagi:variational_1994}
Ichiyanagi, M.: Variational principles of irreversible processes.
\newblock Phys. Rep. \textbf{243}, 125--182 (1994).
\newblock \doi{https://doi.org/10.1016/0370-1573(94)90052-3}

\bibitem{kaufman:dissipative}
Kaufman, A.N.: Dissipative hamiltonian systems: A unifying principle.
\newblock Phys. Lett. A \textbf{100}(8), 419--422 (1984).
\newblock \doi{https://doi.org/10.1016/0375-9601(84)90634-0}

\bibitem{kaufman:algebraic}
Kaufman, A.N., Morrison, P.J.: Algebraic structure of the plasma quasilinear
equations.
\newblock Phys. Lett. A \textbf{88}(8), 405--406 (1982).
\newblock \doi{https://doi.org/10.1016/0375-9601(82)90664-8}

\bibitem{L}
Lanczos, C.: The Variational Principles of Mechanics.
\newblock Dover, Mineola (1986)

\bibitem{larmor:obituary}
Larmor, J.: {Dr.~Edward~John~Routh,~F.R.S.}
\newblock Nature \textbf{76}, 200--202 (1907).
\newblock \doi{https://doi.org/10.1038/076200b0}

\bibitem{lemons:perfect}
Lemons, D.S.: Perfect Form: {V}ariational Principles, Methods, and Applications
in Elementary Physics.
\newblock Princeton University Press, Princeton (1997)

\bibitem{levi_civita:lezioni}
{Levi~{C}ivita}, T., Amaldi, U.: Lezioni di Meccanica Razionale, vol.~2.
\newblock CompoMat, Rieti, IT (2012).
\newblock Re-edition of the book published in two volumes by Zanichelli in
various editons from 1923 to 1974 (in {I}talian)

\bibitem{li:analytical}
Li, D.: Analytical Thermodynamics.
\newblock Springer, Cham, CH (2022)

\bibitem{liouville:equations}
Liouville, R.: Sur les \'equations de la dynamique.
\newblock Compt. R. Acad. Sci. Paris \textbf{114}, 1171--1172 (1892).
\newblock \url{https://gallica.bnf.fr/ark:/12148/bpt6k3070h/f1171.item}

\bibitem{lutzen:dissipative}
L\"{u}tzen, J.: Mechanistic Images in Geometric Form. {H}einrich {H}ertz’s
Principles of Mechanics.
\newblock Oxford University Press, Oxford (2005)

\bibitem{machlup:fluctuations}
Machlup, S., Onsager, L.: Fluctuations and irreversible process. {II}.
{S}ystems with kinetic energy.
\newblock Phys. Rev. \textbf{91}, 1512--1515 (1953).
\newblock \doi{10.1103/PhysRev.91.1512}

\bibitem{morrison:bracket}
Morrison, P.J.: Bracket formulation for irreversible classical fields.
\newblock Phys. Lett. A \textbf{100}(8), 423--427 (1984).
\newblock \doi{https://doi.org/10.1016/0375-9601(84)90635-2}

\bibitem{morrison:paradigm}
Morrison, P.J.: A paradigm for joined {H}amiltonian and dissipative systems.
\newblock Physica D \textbf{18}(1), 410--419 (1986).
\newblock \doi{https://doi.org/10.1016/0167-2789(86)90209-5}

\bibitem{mueller:entropy}
M\"uller, I., Weiss, W.: Entropy and Energy. {A} Universal Competition.
\newblock Springer, Berlin (2005)

\bibitem{onsager:reciprocal_I}
Onsager, L.: Reciprocal relations in irreversible processes. {I}.
\newblock Phys. Rev. \textbf{37}, 405--426 (1931).
\newblock \doi{10.1103/PhysRev.37.405}

\bibitem{onsager:reciprocal_II}
Onsager, L.: Reciprocal relations in irreversible processes. {II}.
\newblock Phys. Rev. \textbf{38}, 2265--2279 (1931).
\newblock \doi{10.1103/PhysRev.38.2265}

\bibitem{onsager:fluctuations}
Onsager, L., Machlup, S.: Fluctuations and irreversible processes.
\newblock Phys. Rev. \textbf{91}, 1505--1512 (1953).
\newblock \doi{10.1103/PhysRev.91.1505}

\bibitem{ottinger:dynamics}
\"Ottinger, H.C., Grmela, M.: Dynamics and thermodynamics of complex fluids.
{II}. {I}llustrations of a general formalism.
\newblock Phys. Rev. E \textbf{56}, 6633--6655 (1997).
\newblock \doi{10.1103/PhysRevE.56.6633}

\bibitem{PPGlv}
Podio-Guidugli, P.: A virtual power format for thermomechanics.
\newblock Cont. Mech. Thermodyn. \textbf{20}, 479--487 (2009).
\newblock \doi{10.1007/s00161-009-0093-5}

\bibitem{PPGThD}
Podio-Guidugli, P.: For a statistical interpretation of {H}elmholtz' thermal
displacement.
\newblock Cont. Mech. Thermodyn. \textbf{28}, 1705--1709 (2016).
\newblock \doi{10.1007/s00161-016-0505-2}

\bibitem{SISSA1}
Podio-Guidugli, P.: Continuum Thermodynamics, \emph{SISSA Springer Series},
vol.~1.
\newblock Springer Nature, Cham (2019)

\bibitem{rayle}
Rayleigh, J.W.S.: The Theory of Sound. I \& II.
\newblock Macmillian, London (1877)

\bibitem{routh:treatise}
Routh, E.J.: A Treatise on the Stability of a given State of Motion.
\newblock Macmillan, London (1877).
\newblock Reprinted in \emph{Stability of Motion}, Taylor and Francis, London
1975.

\bibitem{serdyukov:extension}
Serdyukov, S.I., Bel'nov, V.K.: Extension of the variational formulation of the
{O}nsager-{M}achlup theory of fluctuations.
\newblock Phys. Rev. E \textbf{51}, 4190--4195 (1995).
\newblock \doi{10.1103/PhysRevE.51.4190}

\bibitem{sonnet:dissipative}
Sonnet, A.M., Virga, E.G.: Dissipative Ordered Fluids. {T}heories for Liquid
Crystals.
\newblock Springer, London (2012)

\bibitem{strutt}
{Strutt~({L}ord {R}ayleigh)}, J.W.: Some general theorems relating to
vibrations.
\newblock Proc. London Math. Soc. \textbf{4}(1), 357--368 (1873).
\newblock \doi{https://doi.org/10.1112/plms/s1-4.1.357}

\bibitem{raycoll}
{Strutt ({L}ord {R}ayleigh)}, J.W.: Scientific Papers.
\newblock Cambridge University Press, Teddington, England (1883)

\bibitem{swendsen:introduction}
Swendsen, R.H.: An Introduction to Statistical Mechanics and Thermodynamics.
\newblock Oxford University Press, Oxford (2012)

\bibitem{thomson:some_1885}
Thomson, J.J.: {IV}. {O}n some applications of dynamical principles to physical
phenomena.
\newblock Phil. Trans. Roy. Soc. Lond. \textbf{176}, 307--342 (1885).
\newblock \doi{10.1098/rstl.1885.0004}

\bibitem{thomson:some_1887}
Thomson, J.J.: {XVI}. {S}ome applications of dynamical principles to physical
phenomena. {P}art {II}.
\newblock Phil. Trans. Roy. Soc. Lond. A \textbf{178}, 471--526 (1887).
\newblock \doi{10.1098/rsta.1887.0016}

\bibitem{thomson:applications}
Thomson, J.J.: Applications of Dynamics to Physics and Chemistry (lectures of
1886).
\newblock Macmillan, London (1888).
\newblock Reprinted 1968

\bibitem{thomson:treatise}
{Thomson ({L}ord {K}elvin)}, W., Tait, P.G.: Treatise on Natural Philosophy, 2
edn.
\newblock Macmillan, Cambridge (1879).
\newblock Available at
\url{https://books.google.it/books?id=k3lNAAAAYAAJ&printsec=frontcover&hl=it&source=gbs_ge_summary_r&cad=0#v=onepage&q&f=false}

\bibitem{truesdell:rational}
Truesdell, C.: Rational Thermodynamics.
\newblock McGraw-Hill, New York (1969).
\newblock Second edition, Springer, New York 1984.

\bibitem{Tu99}
Tuckerman, M.: On the classical statistical mechanics of non-{H}amiltonian
systems.
\newblock Europhys. Lett. \textbf{45}(2), 149--155 (1999).
\newblock \doi{https://doi.org/10.1209/epl/i1999-00139-0}

\bibitem{Tu}
Tuckerman, M.: Statistical Mechanics: Theory and Molecular Simulation.
\newblock Oxford University Press, Oxford (2010)

\bibitem{TuCic}
Tuckerman, M.E., Liu, Y., Ciccotti, G., Martyna, G.J.: Non-{H}amiltonian
molecular dynamics: {G}eneralizing {H}amiltonian phase space principles to
non-{H}amiltonian systems.
\newblock J. Chem. Phys. \textbf{115}(4), 1678--1702 (2001).
\newblock \doi{10.1063/1.1378321}

\bibitem{van:structure}
V\'an, P., Muschik, W.: Structure of variational principles in nonequilibrium
thermodynamics.
\newblock Phys. Rev. E \textbf{52}, 3584--3590 (1995).
\newblock \doi{10.1103/PhysRevE.52.3584}

\bibitem{virga:rayleigh}
Virga, E.G.: Rayleigh-{L}agrange formalism for classical dissipative systems.
\newblock Phys. Rev. E \textbf{91}, 013203 (2015).
\newblock \doi{10.1103/PhysRevE.91.013203}

\bibitem{wang:onsager}
Wang, H., Qian, T., Xu, X.: Onsager{'}s variational principle in active soft
matter.
\newblock Soft Matter \textbf{17}, 3634--3653 (2021).
\newblock \doi{10.1039/D0SM02076A}

\bibitem{wang:generalized}
Wang, Q.: Generalized {O}nsager principle and its applications.
\newblock In: X.Y. Liu (ed.) Frontiers and Progress of Current Soft Matter
Research, Soft and Biological Matter, chap.~3, pp. 101--132. Springer,
Singapore (2021)

\bibitem{whittaker:treatise}
Whittaker, E.T.: A Treatise on the Analystical Dynamics of Particles and Rigid
Bodies, 4 edn.
\newblock Cambridge University Press, Cambridge (1937).
\newblock Reissued in 1988 and reprinted in 1989.

\end{thebibliography}

%%%%%%%%%%%%%%%%%%
\end{document}